\documentclass[preprint,12pt]{elsarticle}

\graphicspath{ {Images/} }
\usepackage{enumerate}
\usepackage{amsmath}
\usepackage{gensymb}
\usepackage{subcaption}
\usepackage{tabularx}
\usepackage{multicol}
\usepackage{multirow}
\usepackage[normalem]{ulem}
\useunder{\uline}{\ul}{}
\usepackage{longtable}
\usepackage[hidelinks]{hyperref}
\usepackage{booktabs}
\usepackage{tabularx}
\usepackage{rotating}
\usepackage{xcolor}
\usepackage{float}

\usepackage{amssymb}

\begin{document}

\begin{frontmatter}



\title{Renovo: Sensor-Based Visual Assistive Technology for Physiotherapists in the Rehabilitation of Stroke Patients with Upper Limb Motor Impairments}


\author[SSL,IUT]{Mohammad Ridwan Kabir}
\author[NDAG,IUT]{Mohammad Ishrak Abedin}
\author[SSL,IUT]{Mohaimin Ehsan}
\author[NDAG,IUT]{Mohammad Anas Jawad}
\author[SSL,IUT]{Hasan Mahmud}
\author[SSL,IUT]{Md. Kamrul Hasan}

\affiliation[SSL]{organization={Systems and Software Lab (SSL), CSE, IUT}}        
\affiliation[NDAG]{organization={Network and Data Analysis Group (NDAG), CSE, IUT}}

\affiliation[IUT]{organization={Department of Computer Science and Engineering (CSE), Islamic University of Technology (IUT)},
            addressline={Boardbazar}, 
            city={Gazipur},
            postcode={1704}, 
            state={Dhaka},
            country={Bangladesh.}}

\begin{abstract}

Stroke patients with upper limb motor impairments are re-acclimated to their corresponding motor functionalities through therapeutic interventions. Physiotherapists typically assess these functionalities using various qualitative protocols. However, such assessments are often biased and prone to errors, reducing rehabilitation efficacy. Therefore, real-time visualization and quantitative analysis of performance metrics, such as \textit{range of motion}, \textit{repetition rate}, \textit{velocity}, etc., are crucial for accurate progress assessment. This study introduces \textit{Renovo}, a working prototype of a wearable 
motion sensor-based assistive technology that assists physiotherapists with real-time visualization of these metrics. We also propose a novel mathematical framework for generating quantitative performance scores without relying on any machine learning model. We present the results of a three-week pilot study involving 16 stroke patients with upper limb disabilities, evaluated across three successive sessions at one-week intervals by both \textit{Renovo} and physiotherapists (N=5). Results suggest that while the expertise of a physiotherapist is irreplaceable, \textit{Renovo} can assist in the decision-making process by providing valuable quantitative information.

\end{abstract}

\begin{graphicalabstract}
\centering
\includegraphics[width=\textwidth]{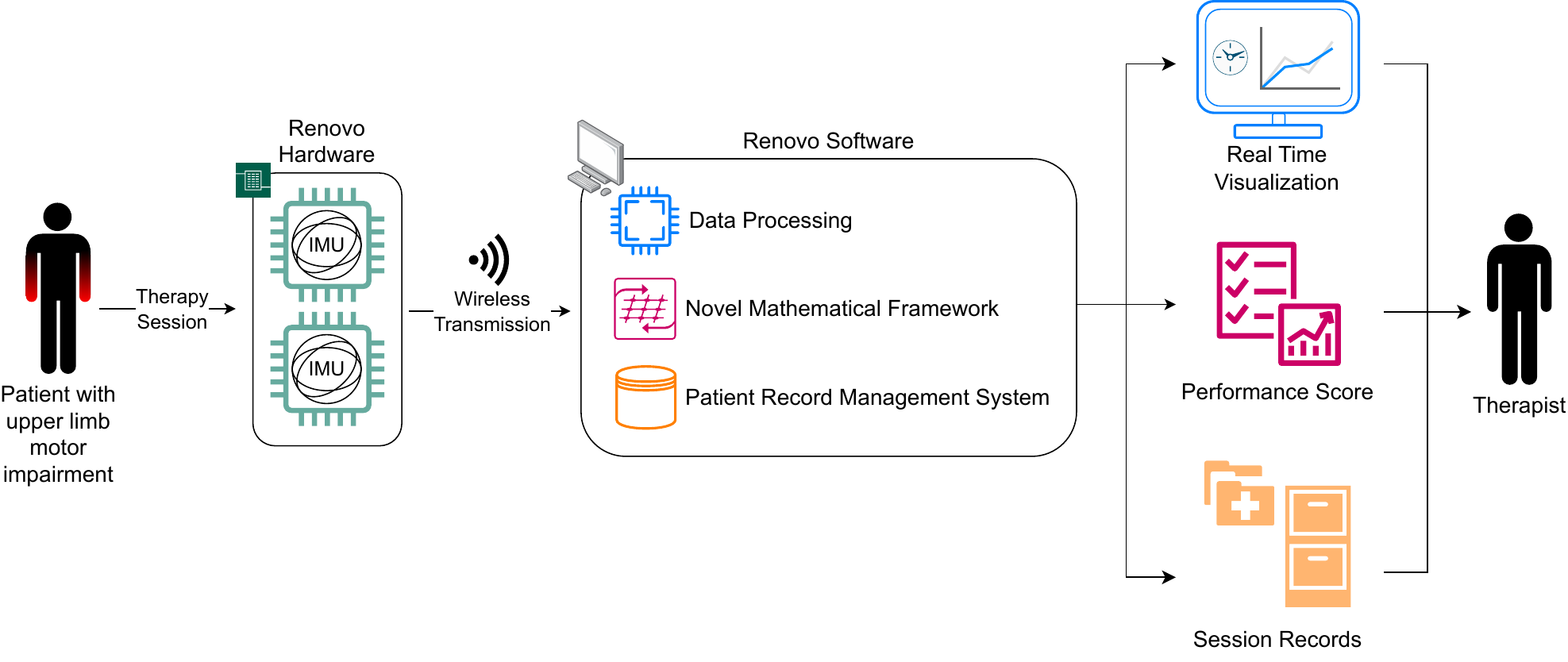}
\end{graphicalabstract}

\begin{highlights}
    \item \textbf{Real-time Visualization and Quantitative Analysis}: This study introduces \textit{Renovo}, a prototype of a sensor-based assistive technology for physiotherapists. It offers real-time visualization and quantitative analysis of performance metrics to accurately assess the progress of stroke patients with upper limb motor impairments.
    
    \item \textbf{Novel Mathematical Framework}: A novel mathematical framework for generating quantitative performance scores in therapeutic interventions without relying on machine learning models is presented. This framework aims to reduce bias and errors in rehabilitation assessments to enhance the efficacy of therapeutic interventions.
    
    \item \textbf{Pilot Study}: A three-week pilot study involving 16 stroke patients was conducted. The study evaluated patients across three successive sessions at one-week intervals using both \textit{Renovo} and physiotherapists (N=5). The results indicated that while the expertise of physiotherapists is irreplaceable, \textit{Renovo} provides valuable quantitative information that can assist in the decision-making process.
    
    \item \textbf{Improved Rehabilitation Assessments}: The findings suggest that \textit{Renovo} can enhance rehabilitation assessments by providing unbiased, accurate, and real-time data. This assists physiotherapists in designing and modifying therapeutic interventions based on the evolving nature of the patient's impairments.
    
    \item \textbf{Stakeholder Satisfaction}: The research highlights the potential of \textit{Renovo} to improve stakeholder satisfaction by providing a reliable and efficient tool for monitoring and assessing the rehabilitation progress of stroke patients, thus supporting better clinical outcomes and patient care.
\end{highlights}

\begin{keyword}
Upper Limb Motor Rehabilitation \sep Sensor-Based Assistive Technology \sep Inertial Measurement Units (IMUs) \sep Real-Time Visualization \sep Mathematical Framework \sep Quantitative Performance Assessment

\end{keyword}

\end{frontmatter}


\section{Introduction}

Rehabilitation of stroke survivors with permanent motor impairments of the upper limb \cite{broeks66lankhorst, timmermans2010influence} requires physiotherapists to design and administer relevant therapeutic interventions in repeated sessions \cite{ploderer2016armsleeve,maceira2019wearable, adomavivciene2019influence,gomez2022monitoring}. 
However, these interventions may require modifications based on - (1) the evolving nature of the impairments \cite{fisher1996evolving,smith2002evolving} and (2) the requirement of unique interventions for addressing simultaneous impairments of upper limb segments \cite{fries1993motor}. Therefore, proper performance assessment is imperative to effective intervention, making it an arduous task for the physiotherapists \cite{Raghavan2015UpperLM, hatem2016rehabilitation,li2017motor}. 

Conventionally, physiotherapists, by leveraging their domain expertise, resort to several state-of-the-art motor impairment assessment protocols \cite{coupar2012predictors, zhang2015objective,santisteban2016upper,rahman2022ai,chen2023designing}, such as - the Fugl-Meyer Assessment Scale (FMAS) \cite{gladstone2002fugl,li2017motor,millar2021international,rahman2022ai}, the Functional Ability Scale (FAS) \cite{patel2010novel}, Disabilities of Arm, Shoulder and Hand (DASH) score \cite{beaton2001measuring}, the Stroke Rehabilitation Assessment of Movement (STREAM) \cite{daley1999reliability}, etc.\ to assess the patients' performance. They do so by observing and qualitatively inspecting - muscle spasticity \cite{gladstone2002fugl,li2017motor}, motor capability, perceived pain \cite{beaton2001measuring}, etc. For example, with the FMAS, the physiotherapists rate the \textit{range of motion} of the upper arm in \textit{three} levels, such as - 0-\textit{``only a few degrees''}, 1-\textit{``decreased''}, and 2-\textit{``normal''}, by inspecting muscle spasticity. 
However, performance assessment using these protocols is subjective, which increases the possibility of bias and/or inaccuracies \cite{zhang2015objective,yu2016remote,li2017motor,rahman2022ai,chen2023designing} and causing a patient to suffer from - \textit{unwanted joint pain}, \textit{increased muscle spasticity}, etc. \cite{gladstone2002fugl,lam2016automated,hughes2020developing,le2020stroke}. Rehabilitation in this manner is inefficient, ineffective, and cumbersome \cite{zhang2015objective,ploderer2016armsleeve,li2017motor,rahman2022ai}. 

Studies suggest that quantitative performance assessment is more reliable than its qualitative counterpart \cite{zhang2015objective,bigoni2016does,yu2016remote,li2017motor,dzhalagoniya2018biomechanical}. 
Furthermore, physiotherapists prefer real-time measurement of different performance metrics of therapeutic interventions such as - \textit{range of motion}, \textit{rate of repetition}, \textit{number of repetitions}, etc., with appropriate visualization, higher accuracy, and data collection rate than the traditional clinical tools \cite{nordin2014assessment,zhang2015objective,yu2016remote,lam2016automated,li2017motor,desplenter2019enhancing,baritz2020analysis}. They also prefer quantitative analysis of the recorded data, making it easier to provide feedback on a patient's progress \cite{lam2016automated,desplenter2019enhancing}. 

Considerable strides have been taken towards developing various frameworks for increasing the efficacy of rehabilitating upper limb motor functionality following stroke by assisting physiotherapists with real-time visualization and quantitative performance evaluation \cite{patel2010novel,del2011estimating,wang2014automated,zhang2015objective,yu2016remote,li2017motor,jung2018wearable,repnik2018using,hesam2019improved,boian2002virtual,zollo2011quantitative,kim2012kinematic,turolla2013virtual,anton2015exercise,ploderer2016armsleeve, baritz2020analysis,boian2002virtual,beursgens2011us, markopoulos2011us,turolla2013virtual,jiang2017towards,lee2019learning}. These systems help analyze the upper limb motion-data acquired using - wearable Inertial Measurement Unit (IMU) that consists of accelerometer and gyroscope for capturing the kinematic properties of such motion \cite{sakoe1978dynamic,patel2010novel,beursgens2011us,markopoulos2011us,del2011estimating,zollo2011quantitative,wang2014automated,zhang2015objective,aggarwal2016doctor,yu2016remote,ploderer2016armsleeve,lam2016automated,sapienza2017using,li2017motor,jiang2017towards,aggarwal2017sophy,jung2018wearable,repnik2018using,maceira2019wearable,hesam2019improved,aggarwal2020physiotherapy,rahman2022ai,boukhennoufa2022wearable}, Electromyography (EMG) sensors for analyzing muscle activity \cite{aggarwal2016doctor,li2017motor,jiang2017towards,repnik2018using,rahman2022ai}, pressure sensors for analyzing muscle strength \cite{aggarwal2016doctor,aggarwal2017sophy,aggarwal2020physiotherapy}, Virtual Reality (VR) \cite{boian2002virtual,turolla2013virtual,rahman2022ai}, robot-assisted systems \cite{zollo2011quantitative,kim2012kinematic,nordin2014assessment,baritz2020analysis,rozevink2021homecare,rahman2022ai}, computer vision \cite{anton2015exercise,chen2018upper,lee2019learning,debnath2022review}, etc. 

Therefore, it may be surmised from the above discussion that real-time visualization of performance metrics and quantitative performance evaluation in rehabilitating patients with upper limb disability, despite numerous efforts, is still of interest to the research community. In these studies, patients' performance in different therapeutic interventions was mostly predicted using machine learning models based on the principles of conventional motor functionality assessment scales, such as - the FAS, the FMAS, the ARAT, etc. However, Schwarz et al. \cite{schwarz2019systematic} in their systematic review of the kinematic approach towards motor functionality assessment of the upper limb following stroke, reported that the FMAS was selected as a reference in about 76\% of the reviewed studies. In terms of the type of therapeutic interventions, most of the prior studies \cite{patel2010novel,del2011estimating,beursgens2011us,markopoulos2011us,kim2012kinematic,turolla2013virtual,ploderer2016armsleeve,jung2018wearable,repnik2018using,jung2018wearable,lee2019learning} explored various reaching and manipulating tasks, while only a few \cite{wang2014automated,zhang2015objective,yu2016remote,li2017motor,jiang2017towards,baritz2020analysis,chae2020development,balestra2021automatic,chen2018upper} explored the basic motions of the upper limb, albeit in a limited manner.

\begin{figure}[htbp]
    \centering
    \includegraphics[width=\textwidth]{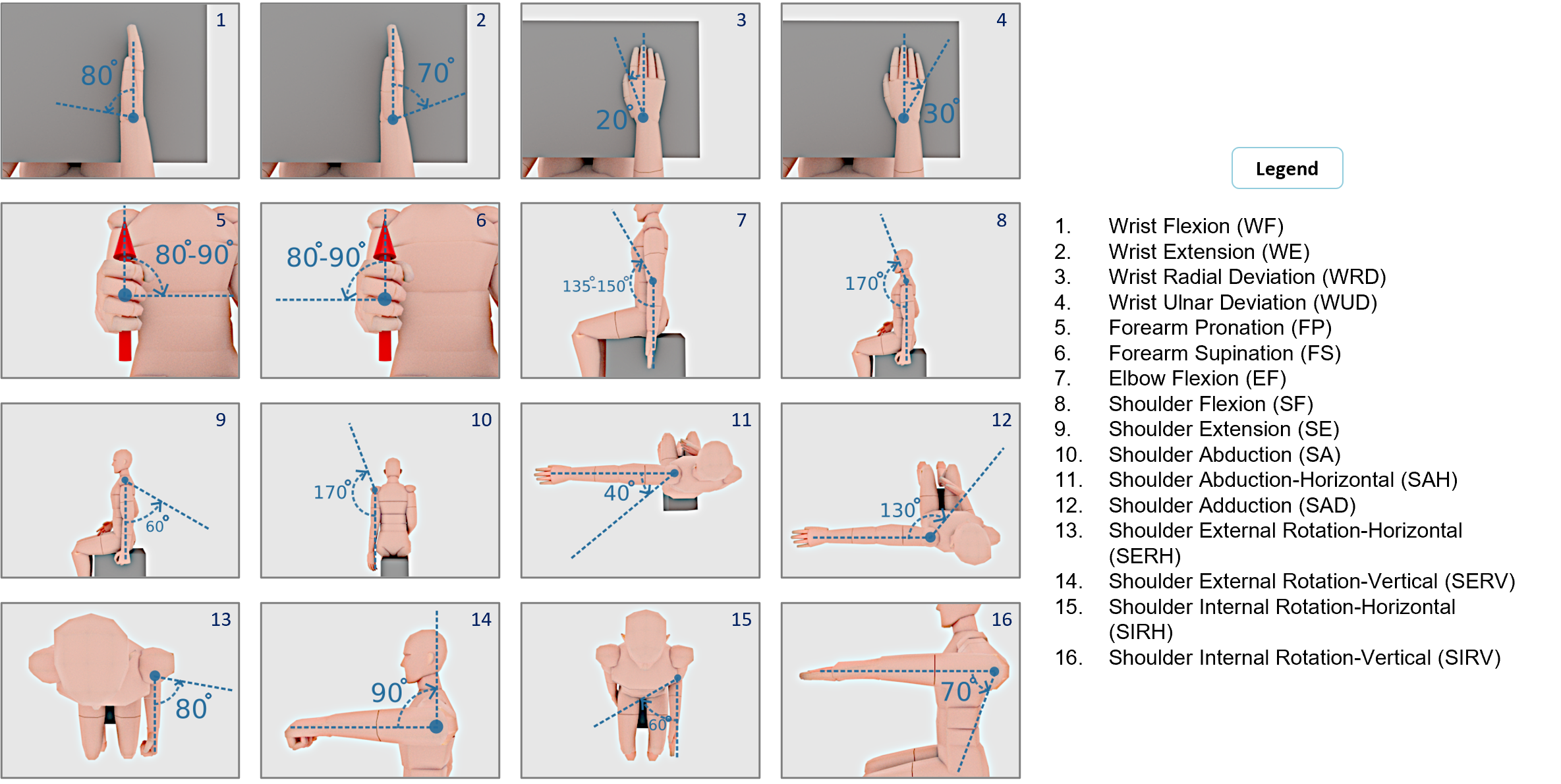}
    \caption{Illustrations of the basic motions of the upper limb \cite{gerhardt2001goniometric,maciejasz2014survey,li2017motor} with respective range of motion \cite{wiki:xxx, vroman2013occupational}.}
    \label{fig:therapies}
\end{figure}

These insights helped identify the scope for developing a wearable IMU-based therapeutic system to assist physiotherapists with various tasks, including \textit{patient management, therapy session organization, real-time visualization of performance metrics, upper limb movement tracking}, etc. while rehabilitating basic arm movements in stroke patients. In this study, we introduce \textit{Renovo}, a prototype of such a system that facilitates the above tasks through a user interface, as shown in \autoref{fig:main interface}, while using the motion data of basic movements of the upper limb transmitted wirelessly by two wearable IMUs ($IMU_1$ and $IMU_2$), as shown in \autoref{fig:sensor placement}. Sixteen of the basic upper limb movements \cite{gerhardt2001goniometric,maciejasz2014survey,li2017motor} such as - \textit{flexion-extension} (wrist, elbow, shoulder), \textit{radial-ulnar deviation} (wrist), \textit{pronation-supination} (forearm), \textit{abduction-adduction} (shoulder), \textit{external-internal rotation} (wrist), etc., as shown in \autoref{fig:therapies} (along with their respective range of motion \cite{wiki:xxx, vroman2013occupational}), were considered in this study. 

Furthermore, we propose a novel mathematical framework that generates a quantitative motor functionality measurement index for performance evaluation during upper limb rehabilitation. In our approach, we reference the FMAS scale considering its widespread use \cite{schwarz2019systematic} and utilize kinematic properties of the upper limb motion data obtained from IMUs without relying on machine learning models to keep the calculation completely explainable. A three-week pilot study was conducted in a clinical setup, with informed consent from 16 stroke patients with impaired upper limb(s) (Mean Age=39.56$\pm$ 16.4 years, 76.92\% Male, 62.5\% with paralyzed left arm) under the supervision of 5 physiotherapists in 3 therapeutic sessions at one-week intervals. Similar to prior studies \cite{patel2010novel,wang2014automated,yu2016remote,li2017motor,sapienza2017using,lee2019learning,adans2020enabling,szturm2021evaluation}, the patients were also evaluated by the physiotherapists in the FMAS scale, which were considered as ground truth values to statistically verify and validate the accuracy and reliability of the system-generated performance scores. Additionally, we conducted a user evaluation of \textit{Renovo} using a paper-based survey with elements from standard questionnaire sets \cite{carayon2009implementation,gil2017useq,anton2018telerehabilitation, dimaguila2019measuring}, separately for the physiotherapists and the patients. In the case of critical decision-making, although there may not be any alternative to the knowledge and expertise of a trained professional, \textit{Renovo} can assist the physiotherapists with the decision-making process by providing meaningful quantitative information. To summarize, the primary contributions of this study are as follows:

\begin{enumerate}[(1)]
    \item Developed a wearable sensor-based therapeutic system to assist physiotherapists with tracking, visualizing, and recording real-time motion data of the upper limbs of stroke patients.
    \item Proposed a novel mathematical framework for quantitative performance assessment of a patient.
\end{enumerate}

In the subsequent section, we will present an in-depth analysis of the existing literature, followed by an introduction to \textit{Renovo} and elaboration on its design, implementation, and workflow. After that, we discuss about our proposed mathematical framework, how we process the motion data to extract features, and the underlying mathematics for generating performance scores using a statistical approach. Then we outline our user study protocol, followed by an analysis of our findings. Finally, we discuss our research challenges, limitations, prospects, and concluding remarks.

\section{Literature Review}
 In the context of this study, the literature may be grouped into three categories - (1) studies focused only on ``\textit{Quantitative Evaluation}'' of motor functionality of the upper limb \cite{patel2010novel,del2011estimating,wang2014automated,zhang2015objective,yu2016remote,li2017motor,jung2018wearable,repnik2018using,hesam2019improved}, (2) studies focused only on ``\textit{Real-time Visualization}'' of various performance metrics of upper limb motion \cite{boian2002virtual,zollo2011quantitative,kim2012kinematic,turolla2013virtual,anton2015exercise,ploderer2016armsleeve, baritz2020analysis}, and (3) studies focused on both \cite{boian2002virtual,beursgens2011us, markopoulos2011us,turolla2013virtual,jiang2017towards,lee2019learning}. In the following sections, we briefly review the relevant literature in these three categories, serving as the basis of our motivation behind this study.

\subsection{Quantitative Evaluation Only}
    Patel et al. \cite{patel2010novel} fed pre-processed data acquired from wearable accelerometers during various reaching and manipulating tasks into a machine-learning model to estimate performance scores using FAS and interpret the progress of motor functionality of impaired upper limb(s) of stroke patients. 
    In a similar study, Li et al. \cite{li2017motor} generated a quantitative evaluation index of upper limb mobility based on the FMAS by analyzing a subset of the basic motions of the upper limb, namely – \textit{flexion-extension} (wrist, elbow, shoulder), \textit{pronation-supination} (forearm), \textit{abduction-adduction} (shoulder), etc. along with few reaching and manipulating tasks, leveraging wearable IMU and EMG sensors. Pre-processed kinematic performance metrics 
    were fed into statistical machine-learning models. They collected data from healthy subjects in similar tasks as a standard performance reference, against which the patients' performances were evaluated. However, in both of these studies, the patients were simultaneously evaluated by physiotherapists using the same scales (FMAS \cite{li2017motor} and FAS \cite{patel2010novel}). The assessment of the physiotherapists was used as ground truth in determining the accuracy of the estimated performance score by the frameworks. Consequently, the system evaluation assisted physiotherapists with customizing the rehabilitation regimen of the patient based on their motor capability. 

    On the other hand, Jung et al. \cite{jung2018wearable}, by employing supervised machine learning algorithms on the motion data acquired from \textit{five} wearable IMUs, developed an intelligent platform that can replicate the motor functionality evaluation skills of a physiotherapist in reaching tasks based on the FMAS. They considered statistical and time-series related features
    to estimate the quality of upper limb motion, while leveraging evaluation metrics, namely - \textit{F-Measure}, \textit{ROC area}, and \textit{RMSE} to measure the quality of the estimation. 

    Repnik et al. \cite{repnik2018using} quantitatively analyzed upper limb movement in reaching and manipulating tasks using IMU and EMG sensors based on the Action Research Arm Test (ARAT) \cite{santisteban2016upper}. Although they quantified certain kinematic parameters of movement, a mathematical basis for a singular motor functionality measurement index was not reported. Zhang et al. \cite{zhang2015objective} developed a similar index based on the FMAS using Dynamic Time Warping (DTW) \cite{sakoe1978dynamic} and machine learning models, with wearable IMU-acquired motion-data only. Similar studies \cite{del2011estimating,wang2014automated,yu2016remote,hesam2019improved} have reliably estimated the performance score in the FMAS by analyzing the kinematic features of IMU-acquired upper limb motion data during motor therapy as well.

\subsection{Real-time Visualization Only}
    
    Ant{\'o}n et al. \cite{anton2015exercise} developed a Kinect-based algorithm that only facilitated real-time visualization and classification of therapeutic interventions on a user-friendly interface. Ploderer et al. \cite{ploderer2016armsleeve} developed ``\textit{ArmSleeve}'' to assist physiotherapists with monitoring the performance of patients with disabilities of the upper limb, leveraging \textit{three} wearable IMUs. 
    The \textit{two} major strengths of ``\textit{ArmSleeve}'', as the authors claim, are - (1) it provides physiotherapists with insights about the daily activities of patients outside therapeutic interventions, and (2) it facilitates communication between physiotherapists and patients regarding the progress of rehabilitation. However, it had its own set of limitations - (1) physiotherapists found it challenging to discern the type of upper limb motions and their correlation to rehabilitation, from the corresponding motion data, and (2) to rate the quality of upper limb movement, physiotherapists had to resort to qualitative inspection based on their subjective knowledge through discussion sessions with patients, despite quantification and visualization of performance metrics.
    As a result, the issue of biased and/or inaccurate assessments remained unresolved.

    Apart from assisting physiotherapists in the rehabilitation of patients with upper limb disability, studies have also explored the same for lower limb disabilities with real-time visualization. In this context, Lam et al. \cite{lam2016automated} developed the ``\textit{Automated Rehabilitation System} (ARS)'' to assist both patients and physiotherapists in the rehabilitation process following hip and knee replacement surgery using wearable IMUs. 
    However, ARS mainly assisted physiotherapists with visualization of performance informatics of patients in respective interventions over time and not with any performance evaluation score. To enhance the efficacy of telerehabilitation of lower limb movements through video consultations, Aggarwal et al. \cite{aggarwal2016doctor,aggarwal2017sophy,aggarwal2020physiotherapy} have developed ``\textit{SoPhy}'', leveraging motion-data acquired from sock-embedded wearable IMUs and pressure sensors. However, although various performance metrics can be visualized in real-time using ``\textit{SoPhy}'', the patients' performance evaluation solely depended on the qualitative inspection by physiotherapists similar to ``\textit{ArmSleeve}'' \cite{ploderer2016armsleeve}.
    
\subsection{Both Quantitative Evaluation and Real-time Visualization}
    Combining both visualization aspects and generation of a quantitative motor functionality assessment index, Lee et al. \cite{lee2019learning} facilitated real-time visualization of - upper limb movement in real-time with a Kinect v2 sensor on a standalone user interface. They adopted a machine learning approach towards quantifying an assessment score, based on the FMAS by analyzing the kinematic properties of upper limb motion. Furthermore, physiotherapists were also recruited to assess the same to evaluate the agreement between their score and the system-generated one, similar to prior studies \cite{patel2010novel,wang2014automated,yu2016remote,li2017motor}. 

    Researchers \cite{beursgens2011us, markopoulos2011us} have also developed ``\textit{Us'em}'', consisting of wearable IMUs embedded in wireless wristband-like and watch-like devices. These devices help monitor the daily movements of the impaired upper limbs of stroke patients while providing them with real-time statistics of the motion.
    Although the quality of the monitored movements was translated into a score, the mathematical basis of such translation was not reported. Consequently, the validity and reliability of these scores press concern. 

    Rozevink et al. developed the ``\textit{hoMEcare aRm rehabiLItatioN}'' (MERLIN) platform that allows stroke patients to rehabilitate their upper limb motor functionality through gamification \cite{rozevink2021homecare}. Their platform facilitated telerehabilitation as well. As a result, physiotherapists could monitor patients' progress and specify game settings remotely. Apart from evaluating patients' progress using the Wolf Motor Function Test (WMFT), they measured their quality of life, user satisfaction, and motivation. In a similar study \cite{szturm2021evaluation}, researchers conducted a feasibility study on game-assisted rehabilitation of upper limb motor functionality with a cohort of 10 stroke patients. Quantitative performance analysis was done using WMFT alongside computerized assessment of specific object manipulation tasks. Their findings reported substantial improvement in patients' performance before and after the rehabilitation program.

    With a similar goal, Jiang et al. \cite{jiang2017towards} developed an Android application for upper limb rehabilitation with real-time visualization of various performance metrics, therapy, and patient management features. They used IMUs, EMG sensors, and temperature sensors to gather upper limb motion data, which after necessary pre-processing, was fed to a machine learning algorithm to quantify patients' motor capability with the FMAS as a reference scale. 

    Promising results have also been obtained using VR in rehabilitating patients with upper limb disabilities \cite{boian2002virtual,turolla2013virtual} highlighting the importance of feedback from performing specific motor tasks using VR devices in such therapies. 
    Wearable robotic exoskeletons with similar objectives \cite{zollo2011quantitative,kim2012kinematic,baritz2020analysis} have also been developed, however, they require very complex and costly setups.\\ 

\section{Renovo: Design and Implementation}
    \textit{Renovo} comprises two entities: (1) a \textit{wearable device}, featuring IMUs and (2) a \textit{user interface}. The \textit{wearable device} registers upper limb motion data and transmits them wirelessly to a \textit{user interface}, running on a host PC, for real-time visualization and performance assessment.
        
    \subsection{The Wearable Device}
\begin{figure}[htbp]
            \centering
            \includegraphics[width=.35\textwidth]{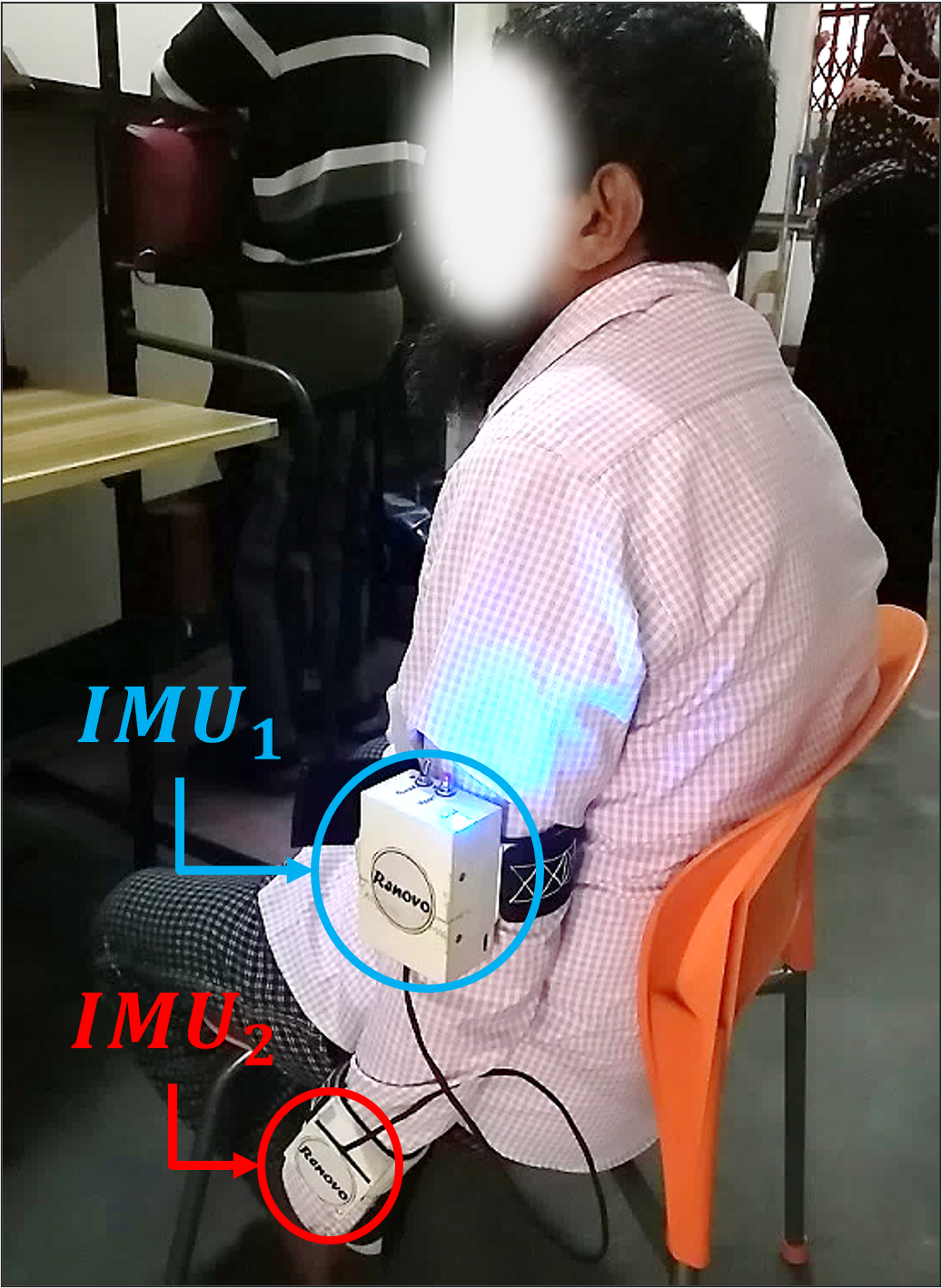}
            \caption{Placement of the wearable sensors ($IMU_1$ and $IMU_2$) on a patient's affected left upper limb. $IMU_1$ is worn on the upper arm, while the position of $IMU_2$ varies depending on the intervention (worn on the forearm in this case).}
            \label{fig:sensor placement}
        \end{figure}

        \textit{Renovo} utilizes \textit{two} IMU sensors to acquire motion data of the upper limb, followed by further processing to generate the corresponding \textit{yaw}, \textit{pitch}, and \textit{roll} motions using an orientation filter \cite{madgwick2010efficient} and wireless transmission to a host PC. This filter prevents the accumulation of angular measurement errors over time while having insignificant ($<5\degree$) instantaneous measurement errors. As shown in \autoref{fig:sensor placement}, one of the two IMUs ($IMU_1$), along with a microcontroller unit and a wireless module, is worn on the patient's upper arm. However, placement of the second IMU ($IMU_2$) varies according to the type of intervention; for example, in wrist and forearm-related exercises, it is placed on the dorsal side of the hand, whereas in Elbow Flexion, it is placed on the forearm. Both of the wearable IMUs were secured with velcro straps in such a way that the \textit{x}-axis of both sensors pointed towards the shoulder joint. 

    \subsection{User Interface}
\begin{figure}[htbp]
            \centering
            \includegraphics[width=\textwidth]{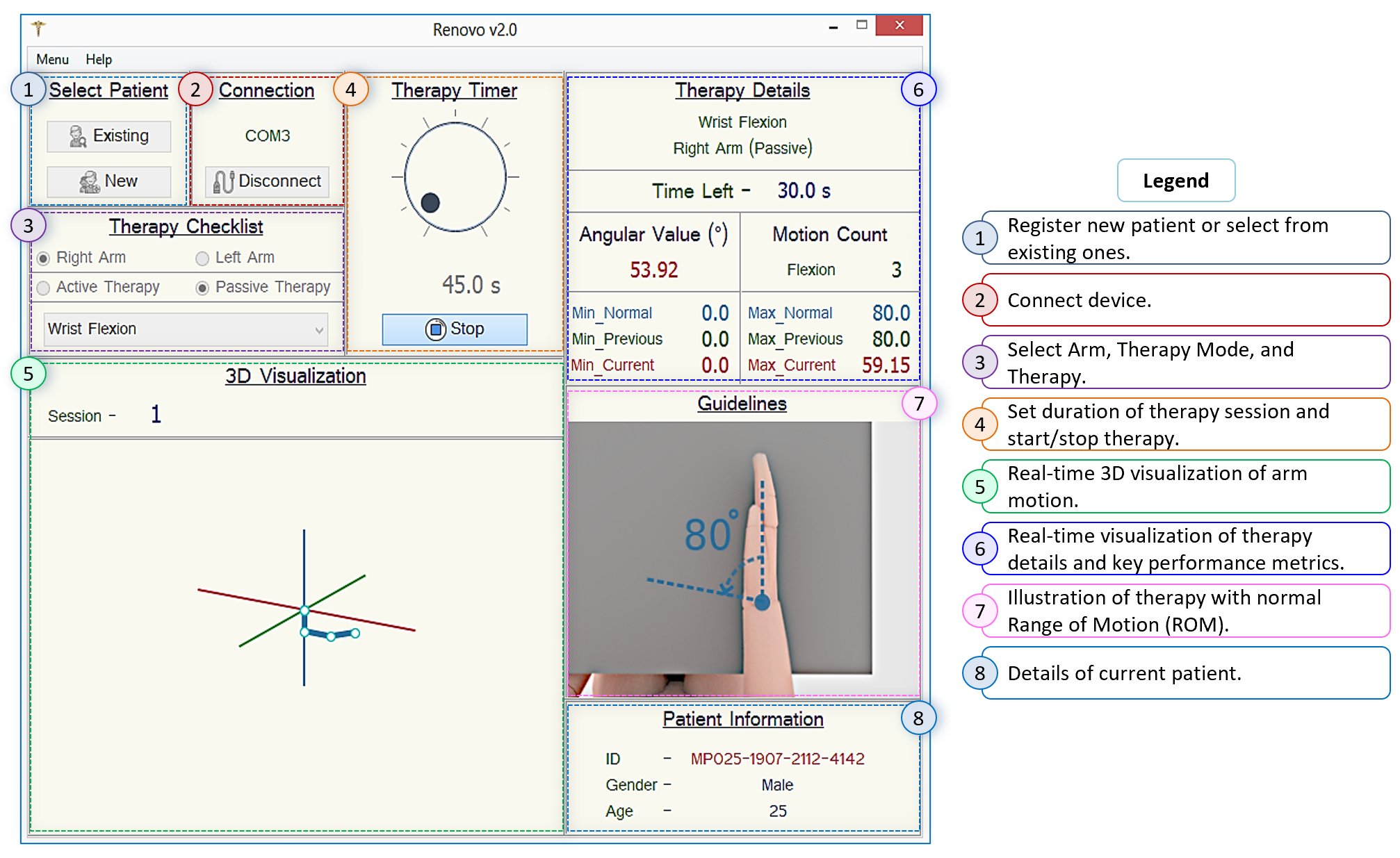}
            \caption{Layout of the \textit{user interface} of \textit{Renovo}.}
            \label{fig:main interface}
        \end{figure}

\begin{figure}[htbp]
            \centering
            \includegraphics[width=\textwidth]{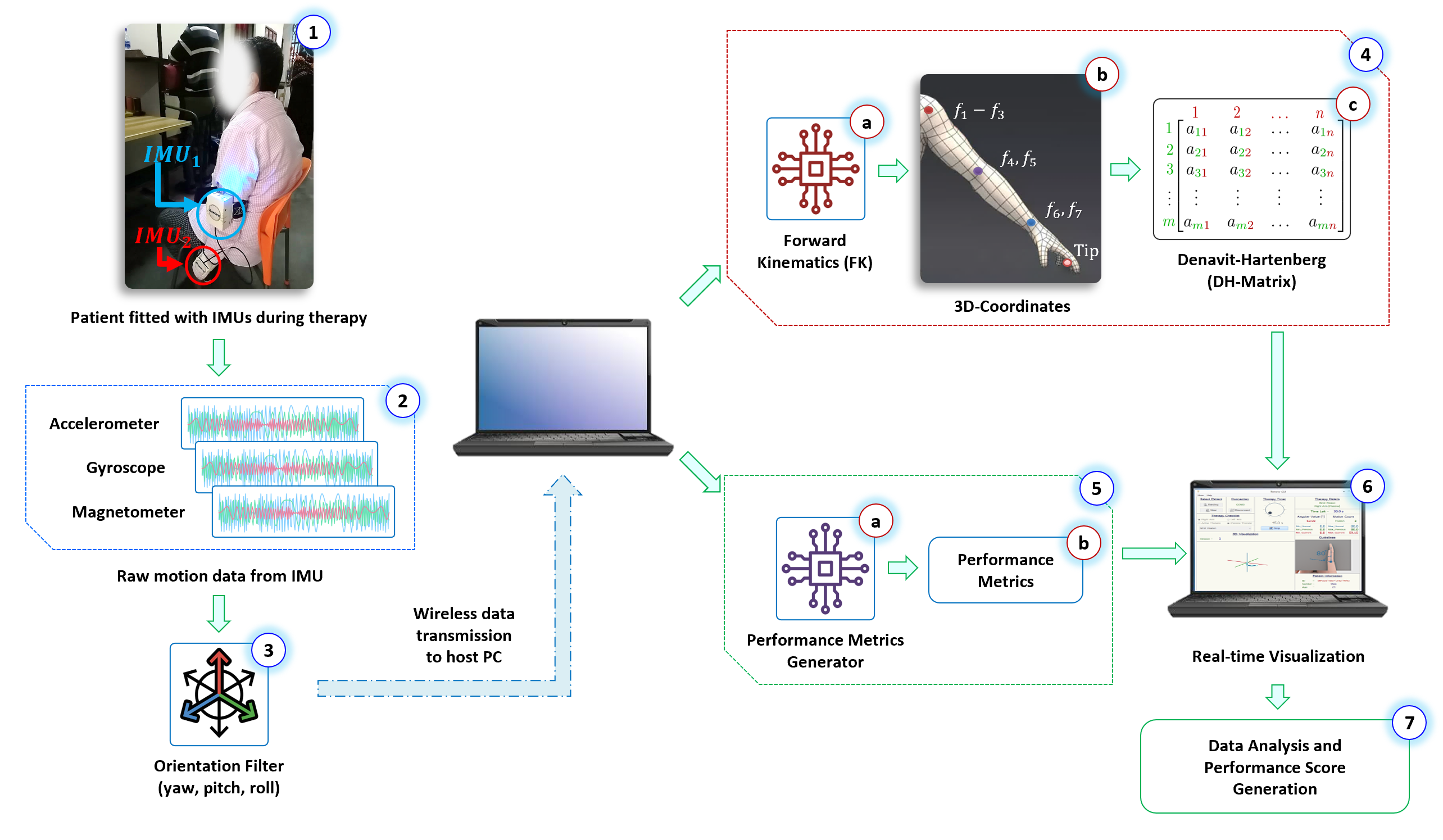}
            \caption{Workflow diagram of \textit{Renovo}. Motion data is acquired from the inertial sensors, followed by data processing and real-time visualization using the \textit{user interface}.}
            \label{fig:renovo workflow}
        \end{figure}

        Desplenter et al. \cite{desplenter2019enhancing} identified visualization as a key factor that makes the task of providing patients with feedback on their progress easier for the physiotherapists. In the same study, the authors identified a few software requirements of such therapeutic systems, some of which include: \textit{automated collection and storage of data from external digital devices}, and \textit{complete numerical analysis and visualizations of the quantitative data}. In light of these requirements, the \textit{user interface} of \textit{Renovo}, as shown in \autoref{fig:main interface}, assists physiotherapists with real-time 3D-tracking of the upper limb motion of a patient in a particular therapy applying forward kinematics \cite{hartenberg1964kinematic} with the Denavit-Hartenberg convention \cite{balasubramanian2011denavit}, along with visualization of various performance metrics, such as – the \textit{range of motion}, \textit{number of repetitions}, \textit{minimum-maximum angular displacement}, etc. Other visual aspects of Renovo include - \textit{the total number of therapeutic sessions administered so far for a particular patient}, \textit{details of the therapy being administered}, and \textit{a corresponding image illustrating the range of motion of therapy and how it should be performed}. Besides visualization, the \textit{user interface} also helps physiotherapists with \textit{patient management}, \textit{therapy session management}, and \textit{data storage}. A workflow diagram of data processing using \textit{Renovo} is depicted in \autoref{fig:renovo workflow}, for better comprehension.
        
\section{Quantitative Performance Evaluation}
    Quantitative performance scores allow physiotherapists to quantify patients' progress alongside their qualitative evaluation \cite{desplenter2019enhancing}, and to provide feedback to patients to motivate them towards rehabilitation. In this section, we explain the mathematics behind calculating these scores. 
    
    \subsection{Generating Performance Metrics Vector (PMV)}
        To quantify a patient's performance in an intervention, performance metrics need to be derived from the corresponding motion data to form a Performance Metrics Vector (PMV). In the literature, PMVs have been generated using performance metrics that are either only \textit{statistical} \cite{patel2010novel,tran2018robust,lucas2019use,bisio2019ehealth,adans2020enabling,chae2020development,balestra2021automatic} or only \textit{time series-related} \cite{lam2016automated,yang2018iot,bobin2018smart} or both \cite{wang2014automated,capela2015feature,mannini2016machine,o2017activity,jung2018wearable,liu2018use,colombo2019sonichand,miller2020comparison,oubre2020estimating,chen2021measuring,meng2021exploration,chen2023designing}. In our study, we have considered both types of performance metrics, where the statistical performance metrics include – \textit{Standard Deviation} (SD), \textit{Mean} (M), \textit{Rate of Repetition} (RR), and \textit{Median of amplitudes above 80\% of max range of motion} (Med$\text{-}$80), and the time series-related ones include - \textit{RMS-value} (RMS), \textit{Wave-Period} (WP), \textit{Wave-Velocity} (WV), and \textit{Wave-Amplitude} (WA). Combining all these metrics, a vector normalized Performance Metrics Vector (PMV), as shown in \autoref{eq:PMV}, was formed for each participant per session of any intervention.
    
        \begin{equation}
            PMV = [SD, M, RR, Med\text{-}80, RMS, WP, WV, WA]
            \label{eq:PMV}
        \end{equation}
                   
    \subsection{Generating Reference PMV (RPMV)}
\begin{figure}[htbp]
            \centering
            \includegraphics[width=\textwidth]{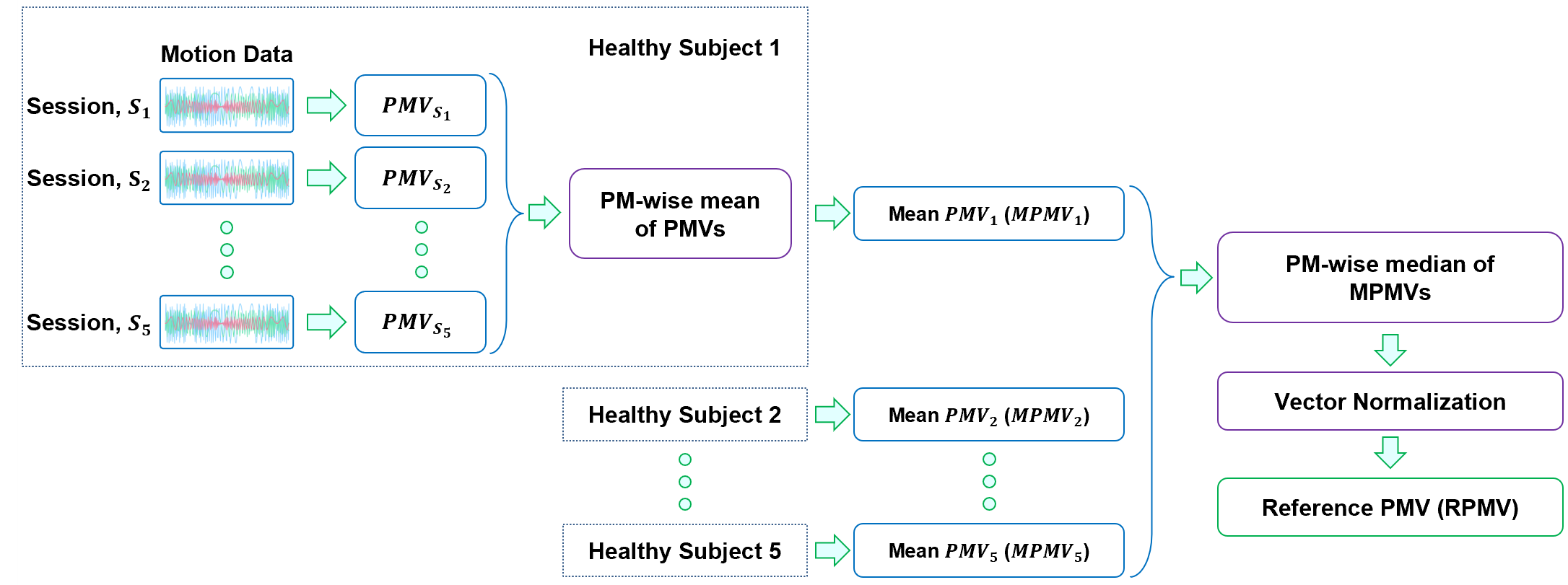}
            \caption{Workflow diagram of generating the Reference PMV (RPMV) of a particular motion of the upper limb from the corresponding data of the 5 healthy subjects.}
            \label{fig:rpmv workflow}
        \end{figure}

        With motivation from prior studies \cite{zhang2015objective,anton2015exercise,li2017motor,repnik2018using,rahman2022ai}, a distance measure was employed to compare the PMV of a motion against its Reference PMV (RPMV) to generate the performance score using our proposed framework. To this end, 5 healthy subjects (Mean Age=24.4$\pm$ 2.4 years, 80\% Male) were recruited to perform 5 consecutive sessions for all the interventions. A PMV was generated for each of the 5 sessions. Afterward, for a particular healthy subject, the performance metric-wise mean of the five PMVs of an intervention was taken to form a Mean PMV (MPMV). In this way, 5 MPMVs were obtained for a particular intervention, from which performance metric-wise median was taken, followed by vector normalization to form the RPMV for that intervention. As a result, 16 RPMVs were obtained, each corresponding to an intervention. For better comprehension, a workflow diagram of generating the RPMV for any of the 16 interventions featured in this study, is depicted in \autoref{fig:rpmv workflow}. 
        
    \subsection{Generating Quantitative Scores}
        To track whether a patient's motor capabilities have improved or relapsed after one session of an intervention, we calculate the \textit{Euclidean Distance} (ED) between the vector-normalized PMV and the RPMV of that intervention for that session. Therefore, for each session of an intervention, a patient's performance has a distance measure (similar to prior studies \cite{zhang2015objective,anton2015exercise,li2017motor,repnik2018using,rahman2022ai}), $\delta_{s \in S}^{\tau \in T}$, where $s$ represents a session out of the total $S$ sessions that have been administered and $\tau$ represents an intervention from the set of all interventions, $T$. For each intervention $\tau$, we form an $\big|S\big| \times \big|S\big|$ Progress Outcome Matrix (POM), $POM^{\tau \in T}$, as shown in \autoref{eq:mat D}, where each element, $POM_{i,j}^{\tau \in T}$, is calculated using \autoref{eq:D}. 
        
        \begin{equation}
                POM_{i,j}^{\tau \in T} = \Delta(s_i, s_j)= \delta_{s_i}^{\tau \in T} - \delta_{s_j}^{\tau \in T}
                \label{eq:D}
        \end{equation}
        
        Since the POM contains the difference between the distance measures for all possible combinations of sessions, the values along the diagonal will be zeros and the remaining entries can either be \textit{less than}, \textit{equal to}, or \textit{greater than} zero. The magnitude of elements of the lower and upper triangular matrices will be equal, with opposite signs. However, we are interested in evaluating the performance of the patients over subsequent sessions. Thus, we need to compute the difference between the distance measures of all possible pairs of subsequent sessions, $s_i$ and $s_j$, where $i > j$. As a result, the total number of elements, $N$, under consideration in the matrix, $POM^{\tau \in T}$, as shown in \autoref{eq:mat D}, is given by \autoref{eq:mat D elements}.
        
        \begin{equation}
            POM^{\tau \in T} = \begin{bmatrix}
                                0 & - & - & \cdots & -\\
                                \Delta(S_2,S_1) & 0 & - & \cdots & -\\
                                \Delta(S_3,S_1) & \Delta(S_3,S_2) & 0 & \cdots & -\\
                                \vdots & \vdots & \vdots & \ddots & \vdots\\
                                \Delta(S_n,S_1) & \Delta(S_n,S_2) & \Delta(S_n,S_3) & \cdots & 0
                            \end{bmatrix}
            \label{eq:mat D}
        \end{equation}
        
        \begin{equation}
            N = \frac{\big|S\big|^2-\big|S\big|}{2}
            \label{eq:mat D elements}
        \end{equation}
        
        Considering a reduced distance measure as a potential indicator of improved performance, elements of the matrix, $POM^{\tau \in T}$, which are \textit{less than}, \textit{equal to}, and \textit{greater than} zero, may be considered as \textit{positive}, \textit{neutral}, and \textit{negative} outcomes, respectively. Therefore, from this matrix, the probability of each of these outcomes, $P(\delta_{s \in S}^{\tau \in T})$, can be calculated using \autoref{eq:prob}, where out of the $N$ elements, $n_p$, $n_n$, and $n_g$ are the counts of \textit{positive}, \textit{neutral}, and \textit{negative} outcomes, respectively. 
        
        \begin{equation}
            P(\delta_{s \in S}^{\tau \in T}) =
            \begin{cases}
                \frac{n_g}{N}, \text{ for } \delta_{s \in S}^{\tau \in T} > 0\\
                \frac{n_n}{N}, \text{ for } \delta_{s \in S}^{\tau \in T} = 0\\
                \frac{n_p}{N}, \text{ for } \delta_{s \in S}^{\tau \in T} < 0
            \end{cases}
            \label{eq:prob}
        \end{equation}
        
\begin{figure}[h]
            \centering
            \includegraphics[width=\textwidth]{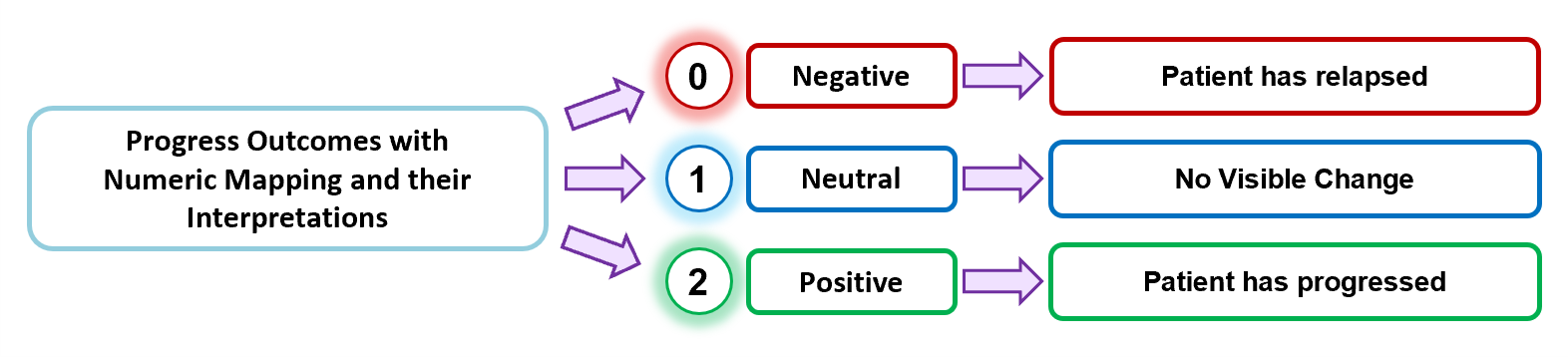}
            \caption{Interpretation of the numerical values ($0, 1$, and $2$) for each of the progress outcomes (\textit{negative, neutral}, and \textit{positive}) in terms of the development of motor functionality.}
            \label{fig:po interpretation}
        \end{figure}

        The FMA scale \cite{gladstone2002fugl} is more effective \cite{rabadi2006comparison} and more commonly used \cite{schwarz2019systematic} for evaluating the rehabilitation progress of a patient with motor impairment(s) of the upper limb(s). It allows physiotherapists to quantify the range of motion of the upper limb as - 0-\textit{``only a few degrees''}, 1-\textit{``decreased''}, and 2-\textit{``normal''} after careful examination of muscle spasticity. Therefore, we have adapted the FMA scale to define the function, $G(\delta_{s \in S}^{\tau \in T})$, as shown in \autoref{eq:score map}, to quantify the progress outcomes of an intervention as \textit{negative}($0$), \textit{neutral}($1$), and \textit{positive}($2$). However, in our case, this mapping is based on quantitative performance analysis rather than subjective evaluation, which resolves the issue of biased and/or inaccurate evaluation \cite{zhang2015objective,yu2016remote,li2017motor,chen2023designing}. The interpretation of the numerical values for each of the progress outcomes in terms of the development of motor functionality is illustrated in \autoref{fig:po interpretation}. 
                    
        \begin{equation}
            G(\delta_{s \in S}^{\tau \in T}) =
            \begin{cases}
                0, \text{ for } \delta_{s \in S}^{\tau \in T} > 0\\
                1, \text{ for } \delta_{s \in S}^{\tau \in T} = 0\\
                2, \text{ for } \delta_{s \in S}^{\tau \in T} < 0
            \end{cases}
            \label{eq:score map}
        \end{equation}
        
        The performance score of a patient in an intervention, $Score^{\tau \in T}$, was generated by calculating the expected value of the outcomes, $E[outcomes]$, as shown in \autoref{eq:score}, ensuring that the maximum achievable score in any particular intervention is 2 following the FMAS scale \cite{gladstone2002fugl}. 
        
        \begin{align}
            Score^{\tau \in T} &= E[outcomes] \nonumber\\
                            &= \sum G(\delta_{s \in S}^{\tau \in T}) \times P(\delta_{s \in S}^{\tau \in T}) \nonumber \\
                          &= 0 \times \frac{n_g}{N} + 1 \times \frac{n_n}{N} + 2 \times \frac{n_p}{N} \nonumber \\
                            &= \frac{n_n + 2n_p}{N}
            \label{eq:score}
        \end{align}
        
\begin{figure}[htbp]
            \centering
            \includegraphics[width=.8\textwidth]{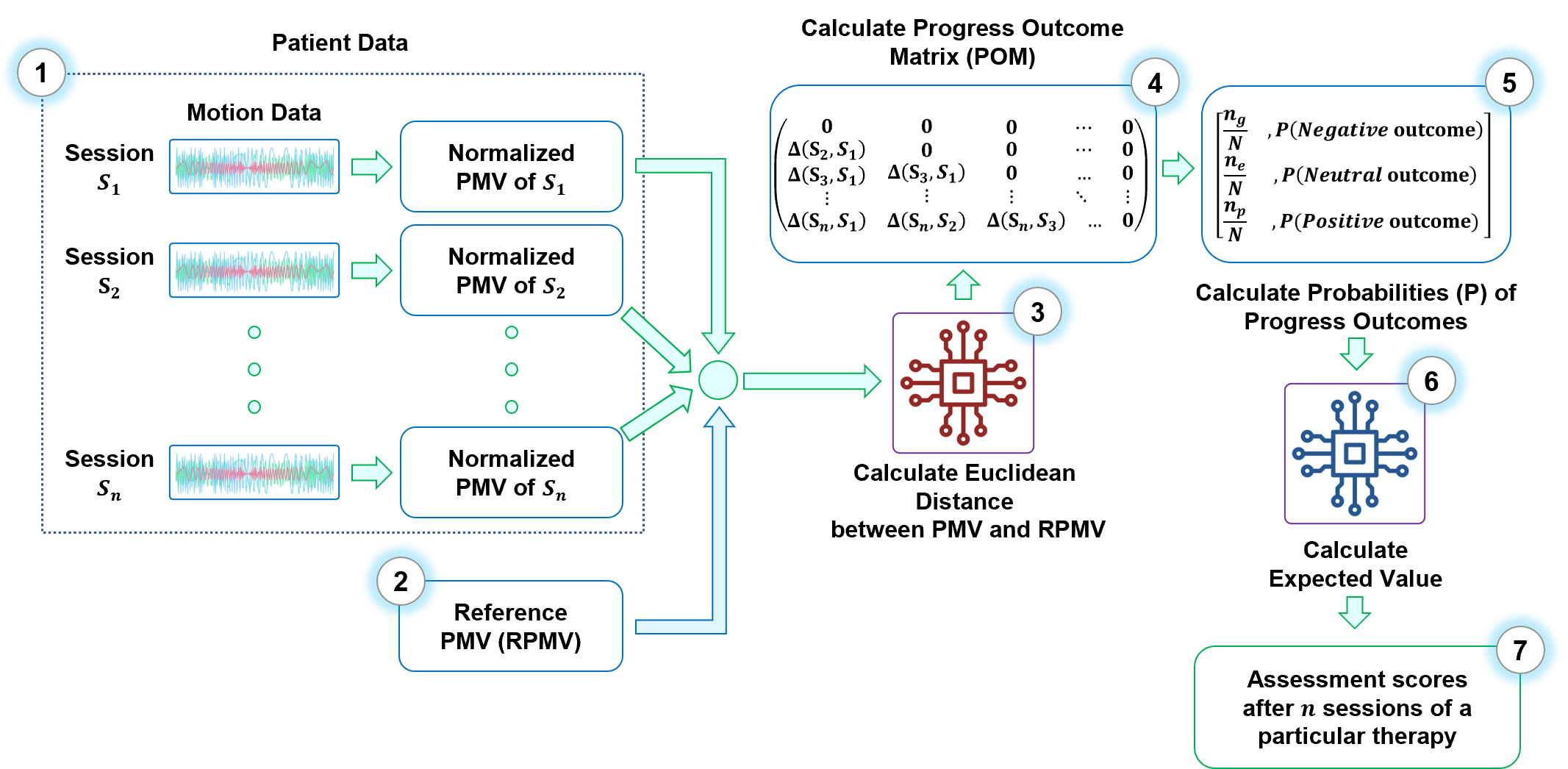}
            \caption{Workflow diagram of generating the performance score of a patient after $n$ sessions of a particular motion of the upper limb, considering the corresponding performance data.}
            \label{fig:performance score}
        \end{figure}

        Thus, the maximum score, $Score_{max}^{\tau \in T}$, achievable by a patient, considering all the 16 interventions, featured in this study, or a subset of it, is given by \autoref{eq:total score}, where $\big |T_{administered}\big |$ is the number of prescribed interventions. For example, if a patient performs 16 interventions, the upper limit of performance evaluation will be 32. For better comprehension, the process of generating the performance score of a patient in a particular intervention is depicted in \autoref{fig:performance score}.

        \begin{equation}
            Score_{max}^{\tau \in T} = 2 \times \big |T_{administered}\big |
            \label{eq:total score}
        \end{equation}
        
    \begin{figure}[H]
        \centering
        \includegraphics[width=0.5\textwidth]{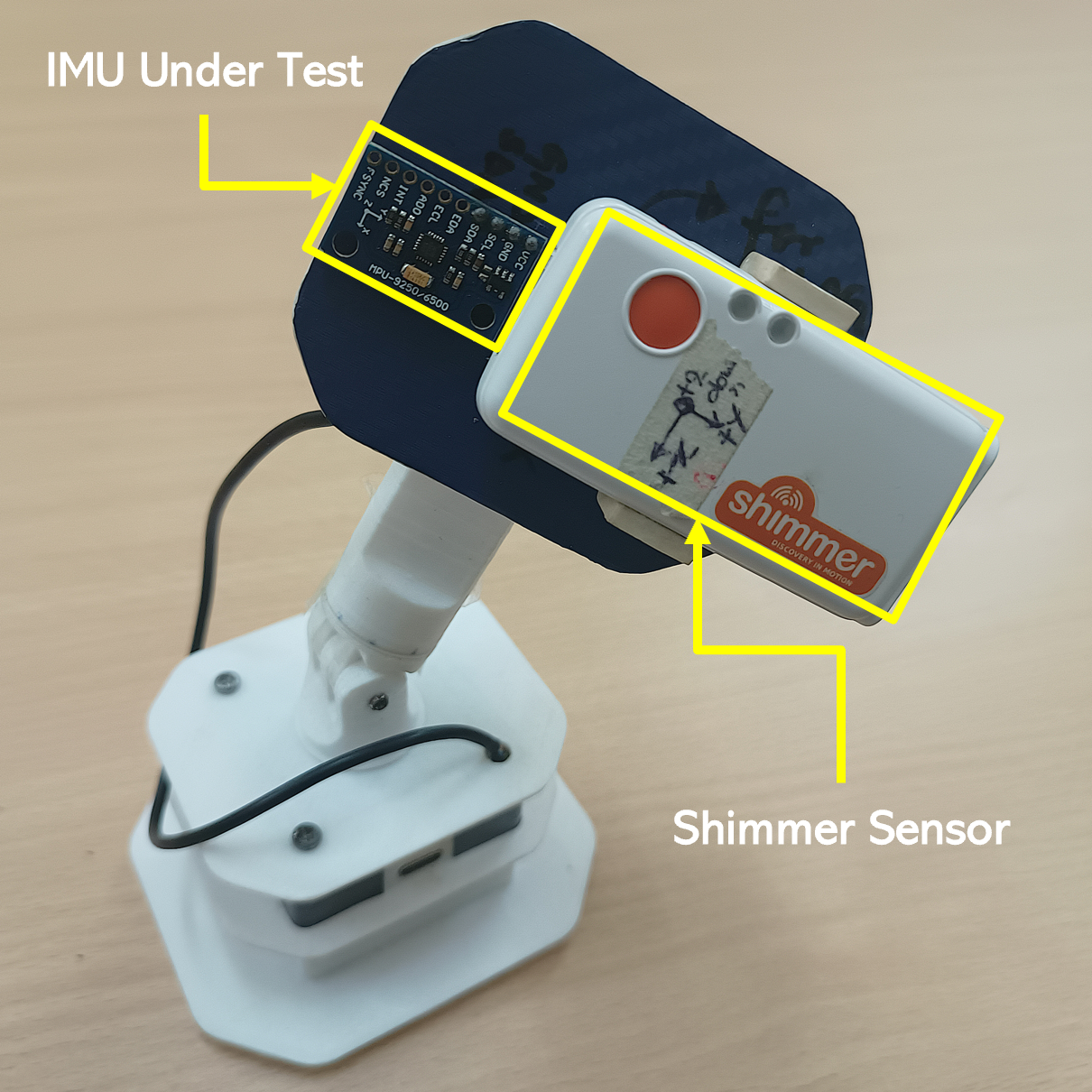}
        \caption{Custom-built rig for comparing sensors.}
        \label{fig:comprig}
    \end{figure}
    \subsection{Sensor Data Validation and Accuracy}
    
    Before we could evaluate our system in a clinical setup, it was necessary to ensure the angular measurements generated by the IMUs used in \textit{Renovo} were valid and trustworthy. Consequently, we compared them against those generated using the IMU of a commercially available Shimmer3 GSR+ sensor \cite{burns2010shimmer,burns2010shimmer2}. In order to carry out the comparison, we built a custom rig capable of simulating ranges of motion in each of the axes, as shown in \autoref{fig:comprig}. Both our device and the Shimmer sensor were mounted on the rig simultaneously and underwent the same set of motions in different axes for two minutes each. The motions were more focused on orientation changes than abrupt acceleration since abrupt accelerations are not common in therapeutic sessions. Moreover, the motions were generated in specific ways, enabling the devices to cover any orientation by combining the orientation of the different axes.

\begin{figure}[htbp]
        \centering
        \includegraphics[width=\textwidth]{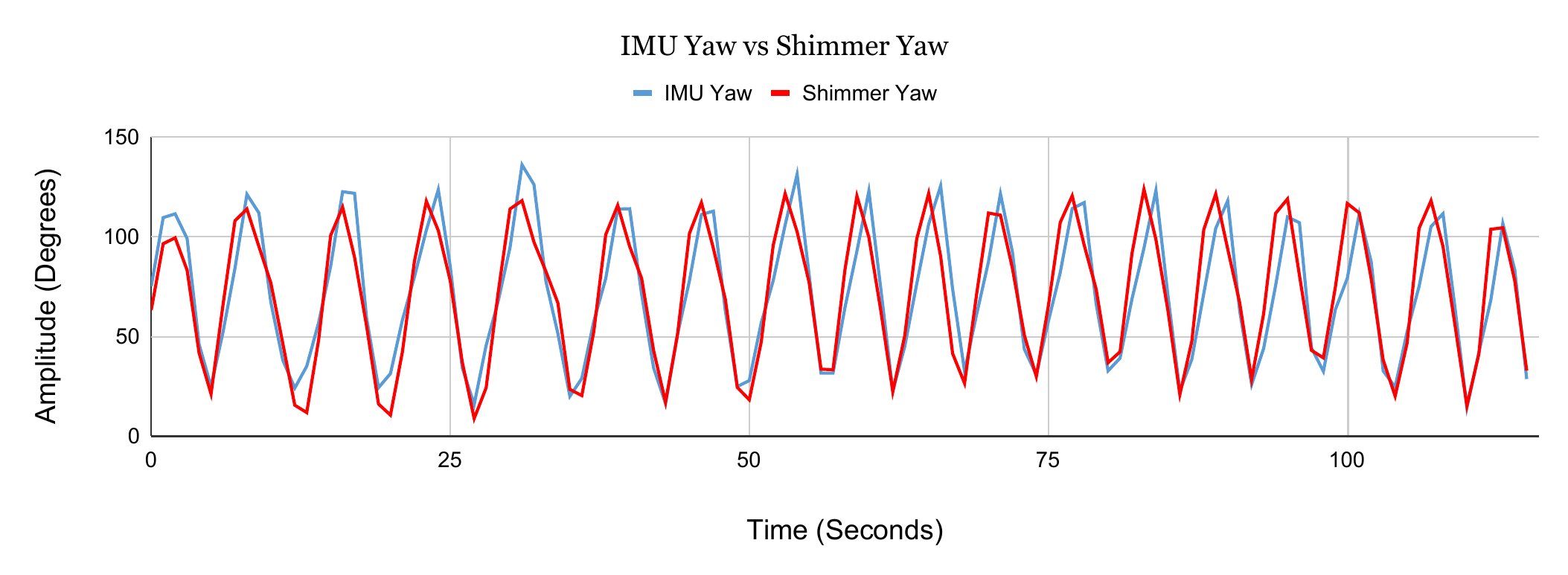}\\
        \includegraphics[width=\textwidth]{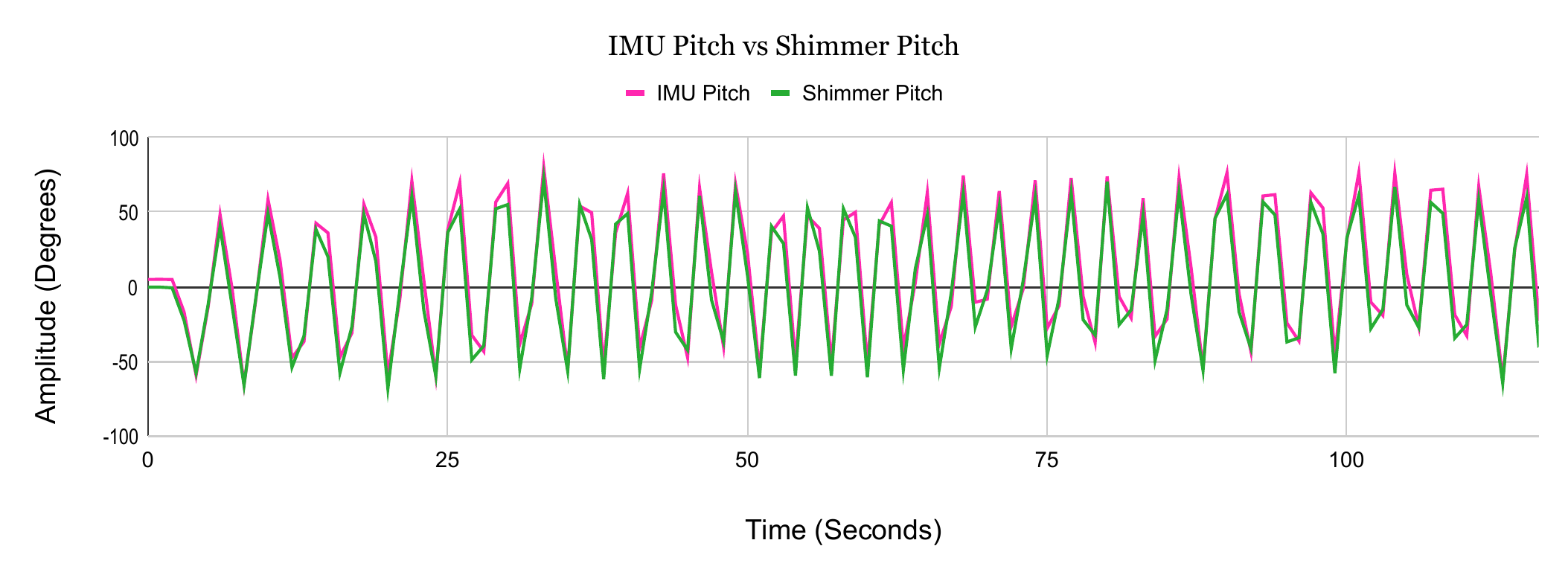}\\
        \includegraphics[width=\textwidth]{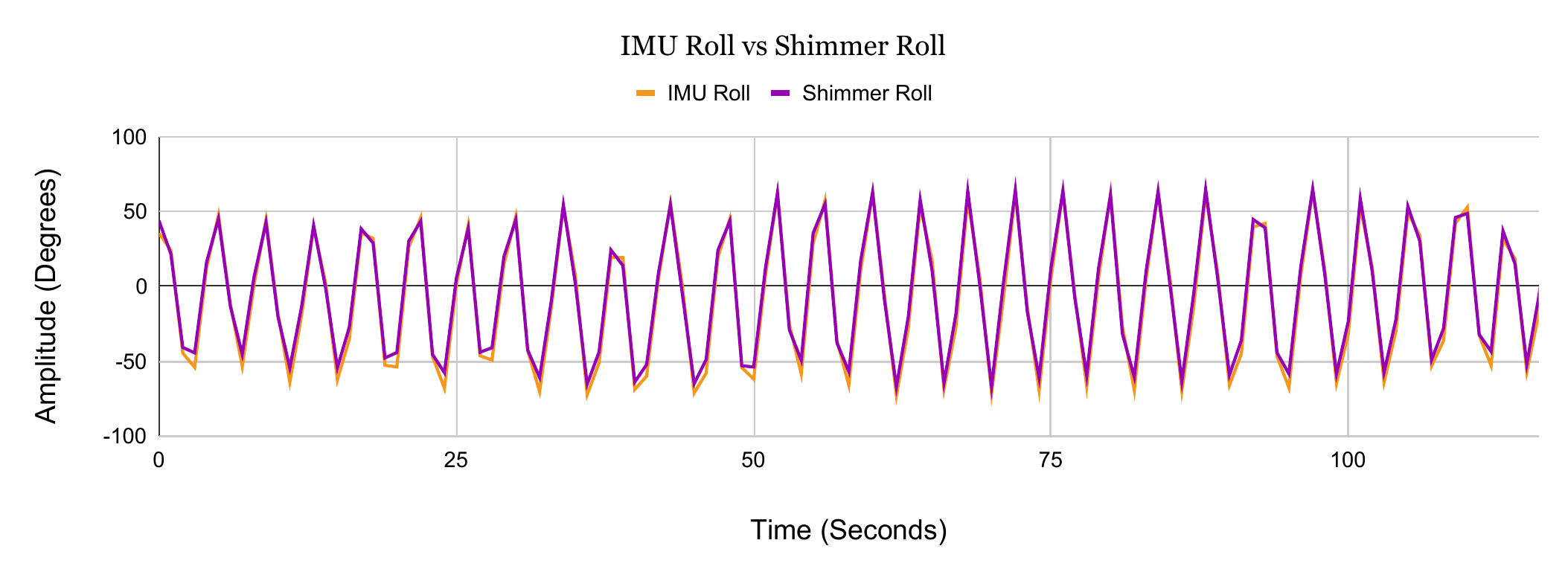}\\
        \caption{Comparison of raw readings of IMU and Shimmer sensor.}
        \label{fig:sensorcomp}
    \end{figure}

    After the motions were completed, the generated data from both of the devices were compared to find out the relative differences. The data were filtered as both of the sensors underwent some initial calibrations and to match the clock timing for both the devices. Following the filtering, the angle over each second was averaged to determine the mean angle in a specific second. Afterwards, the difference between the readings generated by both devices was calculated over each second and then aggregated. The statistics related to the comparison in each of the axes are reported in \autoref{tab:sensorcomp} and the raw visual depiction of the readings is presented in \autoref{fig:sensorcomp}.

    \begin{table}[H]
        \centering
        \caption{Statistical values of sensor comparison.}
        \label{tab:sensorcomp}
        \tiny
        \begin{tabular}{ *{4}{c} }
            \toprule
            \multirow{2}{*}{\textbf{Axis}} & \multicolumn{3}{c}{\textbf{Difference}} \\
            \cmidrule{2-4}
             & \textbf{Mean} & \textbf{Minimum} & \textbf{Maximum} \\
            \midrule
            Yaw & 3.53\% & 0.00\% & 10.40\% \\
            Pitch & 2.68\% & 0.03\% & 6.46\% \\
            Roll & 1.41\% & 0.01\% & 3.45\% \\
            \midrule
            \textbf{Average} & 2.54\% & 0.01\% & 6.77\% \\
            \bottomrule
        \end{tabular}
    \end{table}
    
    It should be noted that the difference between the two sensor readings does not necessarily imply that one sensor is better than the other. Moreover, in the initial phase, discrepancies may exist between the observed values as each sensor may utilize different calibration techniques, leading to higher differences in the observed maximum values. Additionally, since the two sensors utilize different conventions for representing Euler angles, all the yaw values were rotated by specific degrees to report proper directions.
    
    The overall observations suggest that our devices produce fairly accurate sensor data with an average difference of $2.54\%$ across the different axes, which is sufficient for the evaluation task and further validated by the consistent final results tested against a certified professional presented in \autoref{subsec: performance evaluation}.

\section{User Study}
    \subsection{Participants}
        We conducted a \textit{three-week} pilot study with \textit{Renovo} on the rehabilitation of 16 stroke patients with motor impairment(s) of the upper limb(s) (Mean Age=39.56$\pm$ 16.4 years, 76.92\% Male, 62.5\% with paralyzed left arm). All the patients and 5 physiotherapists from two government-recognized rehabilitation centers voluntarily took part in this study with informed consent. The study was approved by the Department of Research, Extension, Advisory, Services, and Publications (REASP) at the Islamic University of Technology (IUT).
        
    \subsection{Experimental Setup}
        Two weeks before the pilot study, an ice-breaking session on the usage details, features, and functionalities of \textit{Renovo}, therapeutic interventions that can be administered with it, and its potential benefits in the rehabilitation of stroke patients with motor impairment(s) of the upper limb(s) was conducted with the physiotherapists. The patients were invited to participate in 3 successive therapeutic sessions (each held at an interval of one week) to perform the interventions as prescribed by their supervising physiotherapists. Each of them was equipped with a wearable device while they performed the interventions while maintaining a sitting posture. In every session, \textit{Renovo} was run on a laptop and placed in front of the patient and the physiotherapists for visual assistance. Due to the limited motor capability of the patients and the development of fatigue, the physiotherapists advised to set the duration of each session to 3 minutes or 180 seconds. Furthermore, the physiotherapists enlightened the patients with the benefits of therapeutic interventions with \textit{Renovo}, while occasionally complimenting them on their performance. This helped the patients build rapport and trust, not only with the physiotherapists but also with the system. At the end of each session, a performance score was generated by the system and stored for future reference. Apart from the system-generated performance score, each patient was concurrently evaluated by all the physiotherapists as well, and an average of these scores was calculated. To ensure the reliability of the system-generated assessment scores, it is necessary to verify that their mean and variance do not differ significantly from that of the physiotherapists. To test for any significant difference in the mean and variance of these scores, we conducted \textit{paired two-tailed t-test for means} and \textit{F-test}, respectively, at a significance level of $\alpha$ =0.05. Furthermore, Pearson’s correlation test was also conducted on these scores to understand the degree of a linear relationship between them. 
        
    \subsection{System Evaluation and Feedback}
        Although the principal users of \textit{Renovo} are the physiotherapists, the patients also benefit as they can visualize their performances and progress in different therapeutic interventions. Therefore, user evaluation and feedback are important to fully comprehend the feasibility and usefulness of the system from the perspective of both the physiotherapists and the patients. In connection to this, we have conducted a paper-based survey, which contained elements from standard questionnaire sets such as the \textit{User Satisfaction Evaluation Questionnaire} (USEQ) \cite{gil2017useq} and \textit{Questionnaire for User Interface Satisfaction} (QUIS) \cite{carayon2009implementation} along with elements from the feedback forms used in different studies \cite{anton2018telerehabilitation, dimaguila2019measuring}. Each participant had to rate the items in the questionnaire on a scale of ``1 - \textit{Very Poor/Strongly Disagree}'' to ``5 - \textit{Excellent/Strongly Agree}''. All the patients and the physiotherapists, who participated in this study, were invited to participate in this survey. Since the system usage of the physiotherapists was different from that of the patients, it is intuitive that their perception of the system would also be different, when different factors are considered. For instance, in the evaluation of user satisfaction with the system, factors such as - \textit{motivation} and \textit{dependence}, are more relevant to the patients, while factors, such as - \textit{success}, \textit{control}, \textit{patient management}, \textit{session management}, \textit{data analysis}, \textit{learnability}, \textit{reliability}, and \textit{assistance}, are more relevant to the physiotherapists. However, some common factors, such as - \textit{experience}, \textit{comfort}, \textit{progress tracking}, \textit{satisfaction}, \textit{acceptability}, \textit{confusion}, \textit{clarity}, \textit{relevance}, and \textit{safety}, were also considered in this regard. Considering the evaluation of users' satisfaction with the system interface, the factors, such as - \textit{aesthetics}, \textit{visualization}, \textit{system alert}, and \textit{performance metrics}, were considered for both the physiotherapists and the patient. Since the physiotherapists are the primary users of the system interface, \textit{two} extra interface-related factors, such as - \textit{terminologies} and \textit{learnability}, were also included. 
        
\section{Results}
    In this section, we analyze the evaluation of patient performance by the system in comparison with that by the physiotherapists, followed by the analysis of user satisfaction with the system and its corresponding interfaces, obtained from the paper-based survey. 
    
    \subsection{Patient Performance Evaluation}\label{subsec: performance evaluation}
        The patient assessment scores by the system and the average of the same by the physiotherapists after 3 successive sessions at one-week intervals are summarized in \autoref{tab:sys_PT scores}, where interventions marked by a ``-'' were not administered. Each intervention was evaluated out of 2 and the maximum total score achievable by each patient varied depending on the number of interventions administered. 

        \begin{sidewaystable}[htbp]
        	\caption{Patients' performance evaluation scores using \textit{Renovo}.}
        	\label{tab:sys_PT scores}
        	\tiny
        	\begin{tabularx}{.88\textwidth}{cccccccccccccccccc}
        		\toprule
        		\textbf{Therapy$^a$} & \textbf{S/P$^b$}
        		 & \textbf{p1} & \textbf{p2} & \textbf{p3} & \textbf{p4} & \textbf{p5} & \textbf{p6}
        		 & \textbf{p7} & \textbf{p8} & \textbf{p9} & \textbf{p10} & \textbf{p11} & \textbf{p12}
        		 & \textbf{p13} & \textbf{p14} & \textbf{p15} & \textbf{p16}\\
        		\midrule

                \multirow{2}{*}{\rotatebox[origin=c]{0}{WF}} &S&1.67&-&-&1.0&0.0&1.0&1.33&1.67&-&1.0&0.33&-&0.0&1.67&1.67&1.33\\
                &P&1.8&-&-&1.4&0.2&0.8&1&1.8&-&1&1.2&-&0.2&1.2&1.6&1.4\\
                
        		\multirow{2}{*}{\rotatebox[origin=c]{0}{WE}} &S&1.33&-&-&-&-&-&0.33&1.0&-&1.0&0.67&-&1.33&0.67&1.0&1.0\\
                &P&1.6&-&-&-&-&-&0.4&1&-&1.2&1&-&1&1.2&1&1\\
          
        		\multirow{2}{*}{\rotatebox[origin=c]{0}{WRD}} &S&1.0&-&-&-&2.0&-&0.67&1.33&-&1.0&1.0&-&-&-&2.0&2.0\\
                &P&1&-&-&-&1.8&-&1&1&-&1&1&-&-&-&1.6&1.8\\
        		
                \multirow{2}{*}{\rotatebox[origin=c]{0}{WUD}} &S&1.67&-&-&-&-&-&0.67&1.0&-&2.0&0.33&-&-&-&0.67&-\\
                &P&1.6&-&-&-&-&-&1&1&-&2&0.8&-&-&-&0.6&-\\
        
        		\multirow{2}{*}{\rotatebox[origin=c]{0}{FP}} &S&-&-&-&2.0&2.0&-&1.67&0.33&-&1.33&-&2.0&0.67&-&1.33&0.67\\
                &P&-&-&-&1.8&2&-&1.4&0.2&-&1&-&2&1&-&1.6&0.8\\
          
                \multirow{2}{*}{\rotatebox[origin=c]{0}{FS}} &S&-&-&-&2.0&2.0&-&0.67&0.33&-&1.67&-&1.0&0.0&-&1.0&1.0\\
                &P&-&-&-&1.6&2&-&1.2&0.2&-&2&-&1&0.2&-&1&1\\
        
    			\multirow{2}{*}{\rotatebox[origin=c]{0}{EF}} &S&1.67&2.0&1.0&0.0&1.0&-&1.33&1.33&-&-&-&-&2.0&-&-&-\\
                &P&1.2&1.8&1&0.8&0.8&-&1.2&1&-&-&-&-&2&-&-&-\\
        
        		\multirow{2}{*}{\rotatebox[origin=c]{0}{SF}} &S&1.0&0.0&-&-&2.0&-&0.67&1.0&-&-&-&-&-&-&-&-\\
                &P&1&0.6&-&-&1.8&-&1&1&-&-&-&-&-&-&-&-\\
          
        		\multirow{2}{*}{\rotatebox[origin=c]{0}{SE}} &S&-&0.0&-&-&-&-&1.33&1.67&-&-&-&-&-&-&-&-\\
                &P&-&0.6&-&-&-&-&1&2&-&-&-&-&-&-&-&-\\
          
        		\multirow{2}{*}{\rotatebox[origin=c]{0}{SA}} &S&-&-&0.0&2.0&2.0&-&-&1.67&-&-&-&-&-&-&-&-\\
                &P&-&-&0.6&2&1.8&-&-&1.8&-&-&-&-&-&-&-&-\\
          
        		\multirow{2}{*}{\rotatebox[origin=c]{0}{SAH}} &S&-&-&-&-&-&-&-&-&-&0.67&-&-&-&-&-&-\\
                &P&-&-&-&-&-&-&-&-&-&1&-&-&-&-&-&-\\
          
        		\multirow{2}{*}{\rotatebox[origin=c]{0}{SAD}} &S&-&-&-&-&-&-&-&-&-&-&-&-&-&-&-&-\\
                &P&-&-&-&-&-&-&-&-&-&-&-&-&-&-&-&-\\
          
        		\multirow{2}{*}{\rotatebox[origin=c]{0}{SERH}} &S&-&-&-&-&-&-&1.0&-&0.67&-&-&0.33&-&-&-&-\\
                &P&-&-&-&-&-&-&1&-&1.2&-&-&0&-&-&-&-\\
          
        		\multirow{2}{*}{\rotatebox[origin=c]{0}{SERV}} &S&-&-&-&-&-&-&-&-&-&0.0&-&1.67&-&-&-&-\\
                &P&-&-&-&-&-&-&-&-&-&0.4&-&2&-&-&-&-\\
          
        		\multirow{2}{*}{\rotatebox[origin=c]{0}{SIRH}} &S&-&-&2.0&1.0&-&-&0.0&-&2.0&-&1.67&-&-&-&-&-\\
                &P&-&-&1.2&1.4&-&-&0.2&-&1.8&-&1&-&-&-&-&-\\
          
        		\multirow{2}{*}{\rotatebox[origin=c]{0}{SIRV}} &S&-&-&-&-&-&-&-&-&-&1.0&-&0.0&-&-&1.33&2.0\\
                &P&-&-&-&-&-&-&-&-&-&0.6&-&0&-&-&1.4&1.6\\
                \midrule
        		\multirow{2}{*}{\rotatebox[origin=c]{0}{\textbf{Score}}} &S&8.34&2.0&3.0&8.0&11.0&1.0&9.67&11.33&2.67&9.67&4.0&5.0&4.0&2.34&9.0&8.0\\
                &P&8.2&3&2.8&9&10.4&0.8&10.4&11&3&10.2&5&5&4.4&2.4&8.8&7.6\\
        
                \textbf{Max Score} &&12.0&6.0&6.0&12.0&14.0&2.0&22.0&20.0&4.0&18.0&10.0&10.0&10.0&4.0&14.0&12.0\\
        		\bottomrule\\
        		
        		\multicolumn{18}{l}{
        		$^a$ WF = Wrist Flexion, WE = Wrist Extension, WRD = Wrist Radial Deviation, WUD = Wrist Ulnar Deviation,}\\
        		\multicolumn{18}{l}{\ \ \ FP = Forearm Pronation, FS = Forearm Supination, SF = Shoulder Flexion, SE = Shoulder Extension,}\\
        		\multicolumn{18}{l}{\ \ \ SA = Shoulder Abduction, SAH = Shoulder Abduction (Horizontal), SAD = Shoulder Adduction,}\\
        		\multicolumn{18}{l}{\ \ \ SERH = Shoulder External Rotation (Horizontal), SERV = Shoulder External Rotation (Vertical),}\\ 
        		\multicolumn{18}{l}{\ \ \ SIRH = Shoulder Internal Rotation (Horizontal), SIRV = Shoulder Internal Rotation (Vertical).}\\ 
        		\multicolumn{18}{l}{$^b$ S = System Generated Score, P = Mean of Scores by the 5 physiotherapists.}
        	\end{tabularx}
        \end{sidewaystable}
        
        The difference between the mean of system evaluation (Mean=6.19, 95\% CI: [4.52, 7.86], IQR: 6.25) and that of the physiotherapists (Mean=6.38, 95\% CI: [4.75, 8.01], IQR: 3.00) was statistically insignificant, $t$(15)=1.39, $p$=0.184, at a significance level of, $\alpha$=0.05. Their variances did not differ significantly as well, $F$(1, 15)=1.05, $p$=0.460, at a significance level of, $\alpha$=0.05. 
           
        \begin{figure}[htbp]
            \centering
            \begin{subfigure}{0.49\textwidth}
                \centering
                \includegraphics[width=\textwidth]{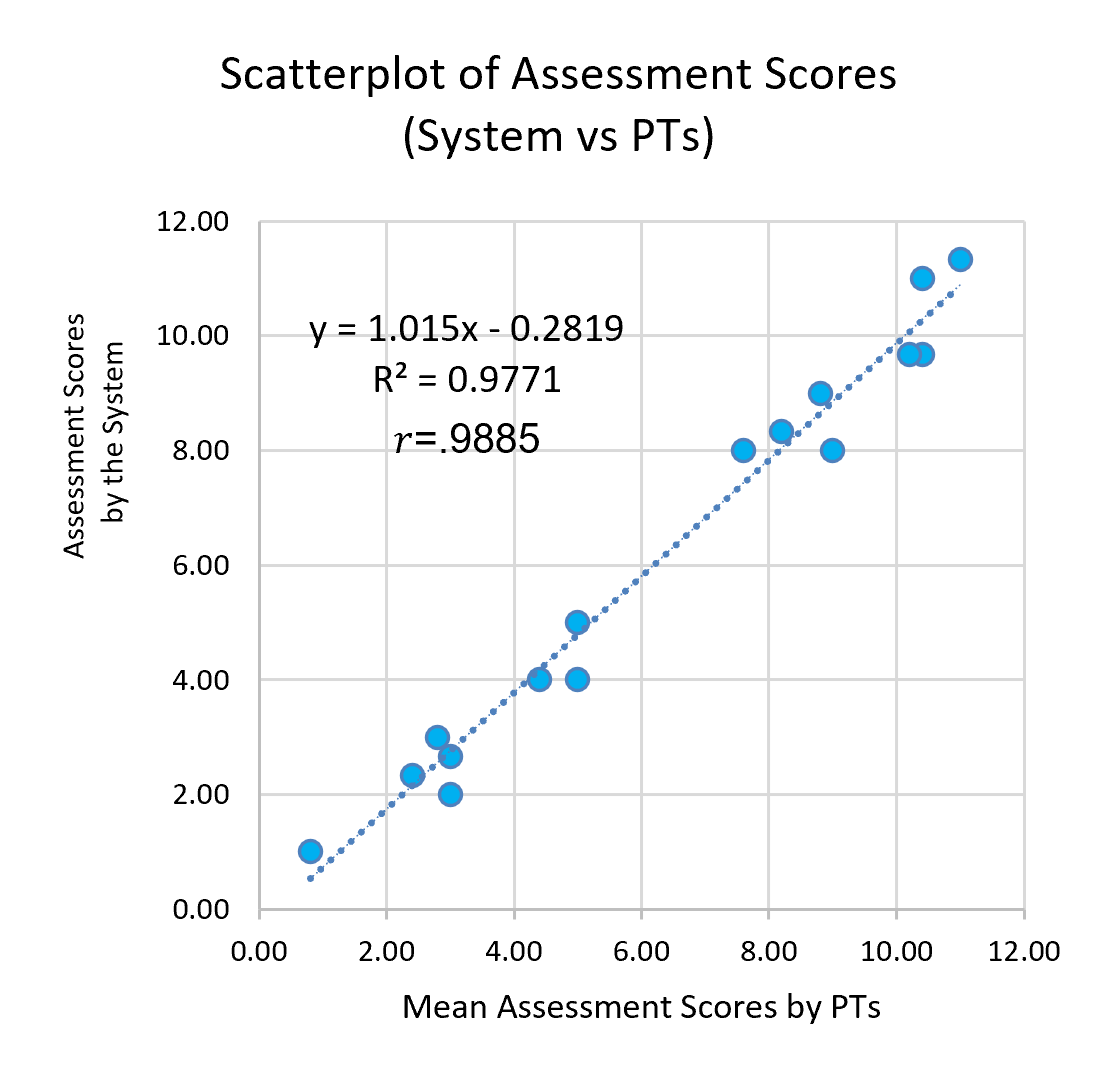}
                \caption{}
                \label{fig:scatter}
            \end{subfigure}
            \hfill
            \begin{subfigure}{0.49\textwidth}
                \centering
                \includegraphics[width=\textwidth]{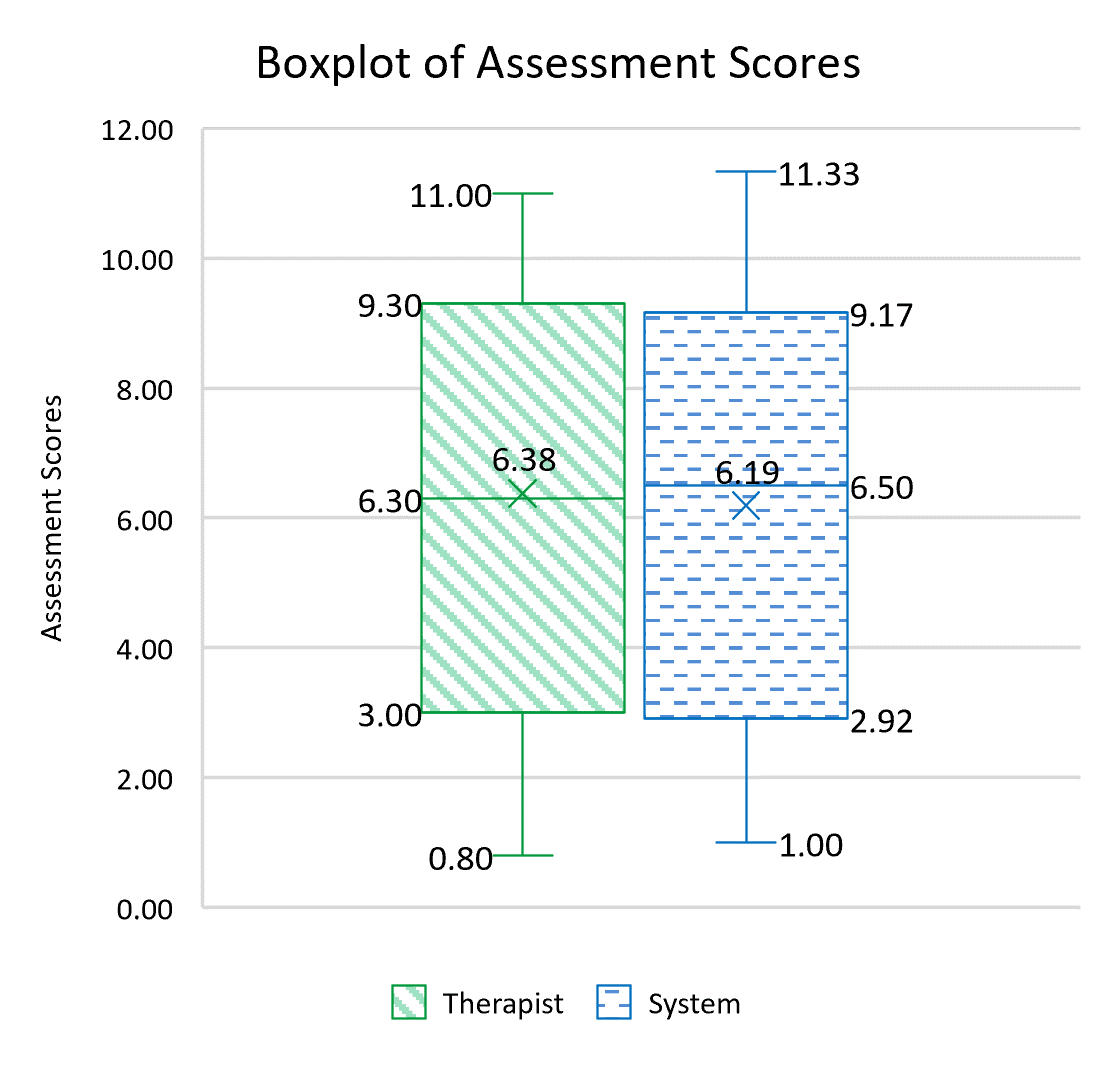}
                \caption{}
                \label{fig:box}
            \end{subfigure}
            
            \caption{(a) Scatter plot of the patient assessment scores by the system and mean of the same by the physiotherapists, showing a strong positive correlation. (b) Box plot of the assessment scores by the system and the physiotherapists.}
            \label{fig:scatter box}
        \end{figure}
        
\begin{figure}[htbp]
            \centering
            \includegraphics[width=\textwidth]{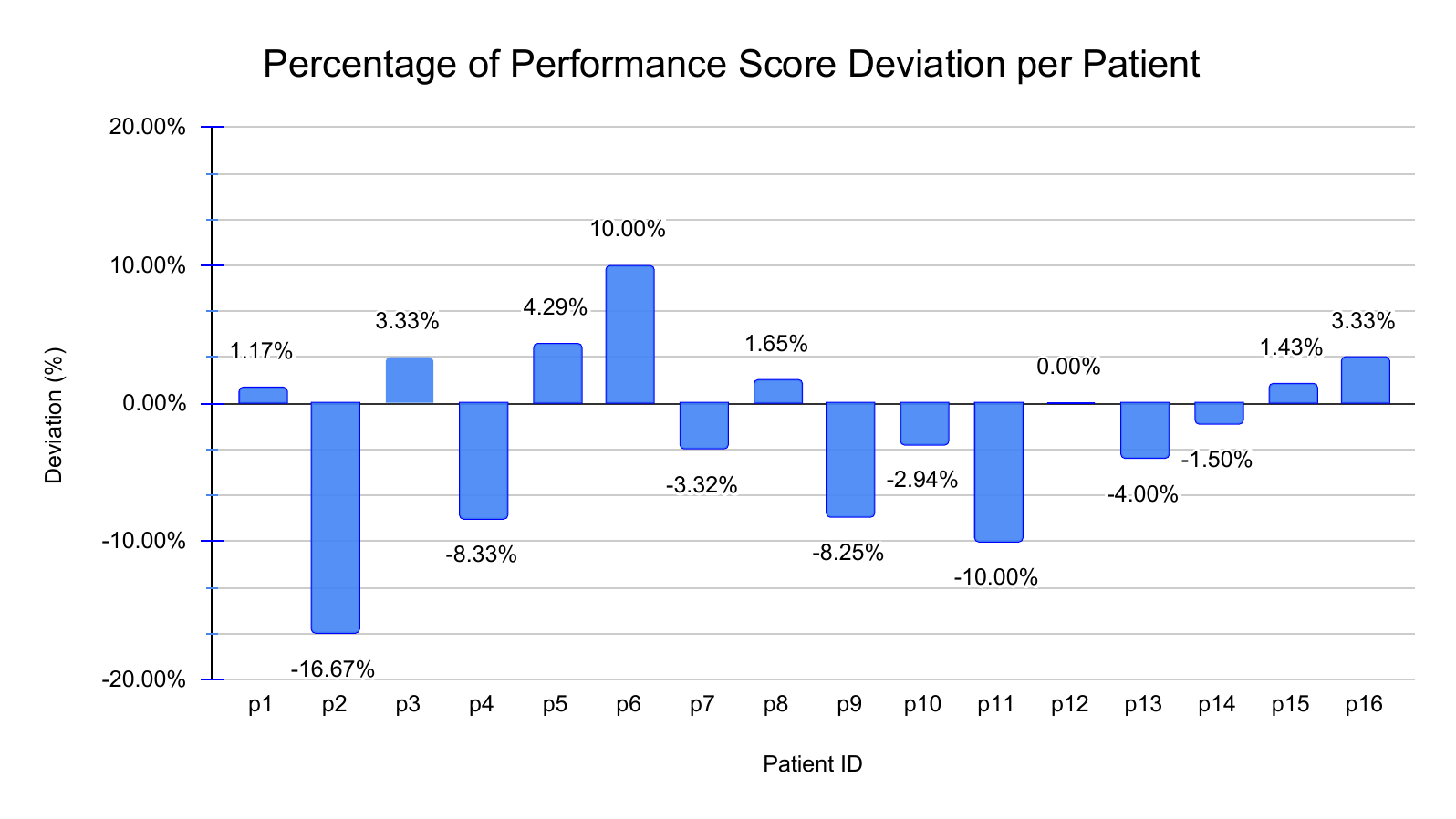}
            \caption{Bar chart of the percentage of deviation of the system-generated patients' performance scores from the same by the physiotherapists.}
            \label{fig:bar}
        \end{figure}

        Similar to prior studies, \cite{yu2016remote,sapienza2017using} we also conducted a regression analysis between the patient evaluation scores by the physiotherapist and that generated using our approach for each patient. The analysis revealed a very good fit ($R^2$ = 0.9771) with a positive correlation of $r$=0.9885. The summary of these scores can be visualized from the scatter plot and the box plot of these scores, as shown in \autoref{fig:scatter} and \autoref{fig:box}, respectively. Furthermore, for each patient, the percentage by which the system-generated scores deviated from the evaluation scores given by the physiotherapists is depicted in \autoref{fig:bar}, where the deviation ranges from a negative 16.67\% to a positive 10.00\%.
        
    \subsection{System Evaluation}
        \begin{table}[htbp]
        	\centering
            \caption{Mean ratings of user satisfaction with \textit{Renovo}.}
        	\label{tab:useq rating summary}
            \tiny
        	\begin{tabularx}{\textwidth}{cccccc}
                \toprule
          
        		\multicolumn{1}{c}{\multirow{2}{*}{\textbf{Factors}}}  & \multicolumn{2}{c}{\textbf{Mean Ratings}* (\textit{out of 5.00})} & \multicolumn{1}{c}{\multirow{2}{*}{\textbf{Factors}}}  & \multicolumn{2}{c}{\textbf{Mean Ratings}* (\textit{out of 5.00})} \\ \cline{2-3} \cline{5-6}\\
                & \textbf{Physiotherapists} & \textbf{Patients}&& \textbf{Physiotherapists} & \textbf{Patients}\\
                \midrule
                Experience &4.40 &4.06 &Data Analysis$ ^{dt}$ &4.40 &-\\
                Comfort &4.80 &4.25 &Session Management$ ^{dt}$ &3.80 &-\\
                Progress Tracking &3.60 &4.50 &Learnability$ ^{dt}$ &4.20 &-\\
                Satisfaction &4.60 &4.50 &Assistance$ ^{dt}$ &4.20 &-\\
                Acceptability &4.00 &4.50 &Reliability$ ^{dt}$ &4.20 &-\\
                Confusion &2.60 &2.63 &Success$ ^{dt}$ &4.00 &-\\
                Clarity &4.40 &4.06 &Patient Management$ ^{dt}$ &3.80 &-\\
                Relevance &4.80 &4.25 &Motivation$ ^{dp}$ &- &4.63\\
                Safety &4.80 &4.88 & Dependence$ ^{dp}$ &- &3.75\\
                Control$ ^{dt}$ &4.00 &- &&&\\\\
                \midrule
                \multicolumn{5}{r}{\textbf{Mean rating across all the factors by Physiotherapists}} &\multicolumn{1}{c}{4.15}\\
                \multicolumn{5}{r}{\textbf{Mean rating across all the factors by Patients}} &\multicolumn{1}{c}{4.18}\\
                \bottomrule\\
                \multicolumn{6}{l}{$ ^*$ Ratings that are marked by a '-', are irrelevant to a particular user (physiotherapist or patient).}\\
                \multicolumn{6}{l}{$ ^{dt}$ Factors relevant to the physiotherapist only.}\\
                \multicolumn{6}{l}{$ ^{dp}$ Factors relevant to the patient only.}
        	\end{tabularx}
        \end{table}
        
\begin{figure}[htbp]
            \centering
            \begin{subfigure}{.95\textwidth}
                \centering
                \includegraphics[width=\textwidth]{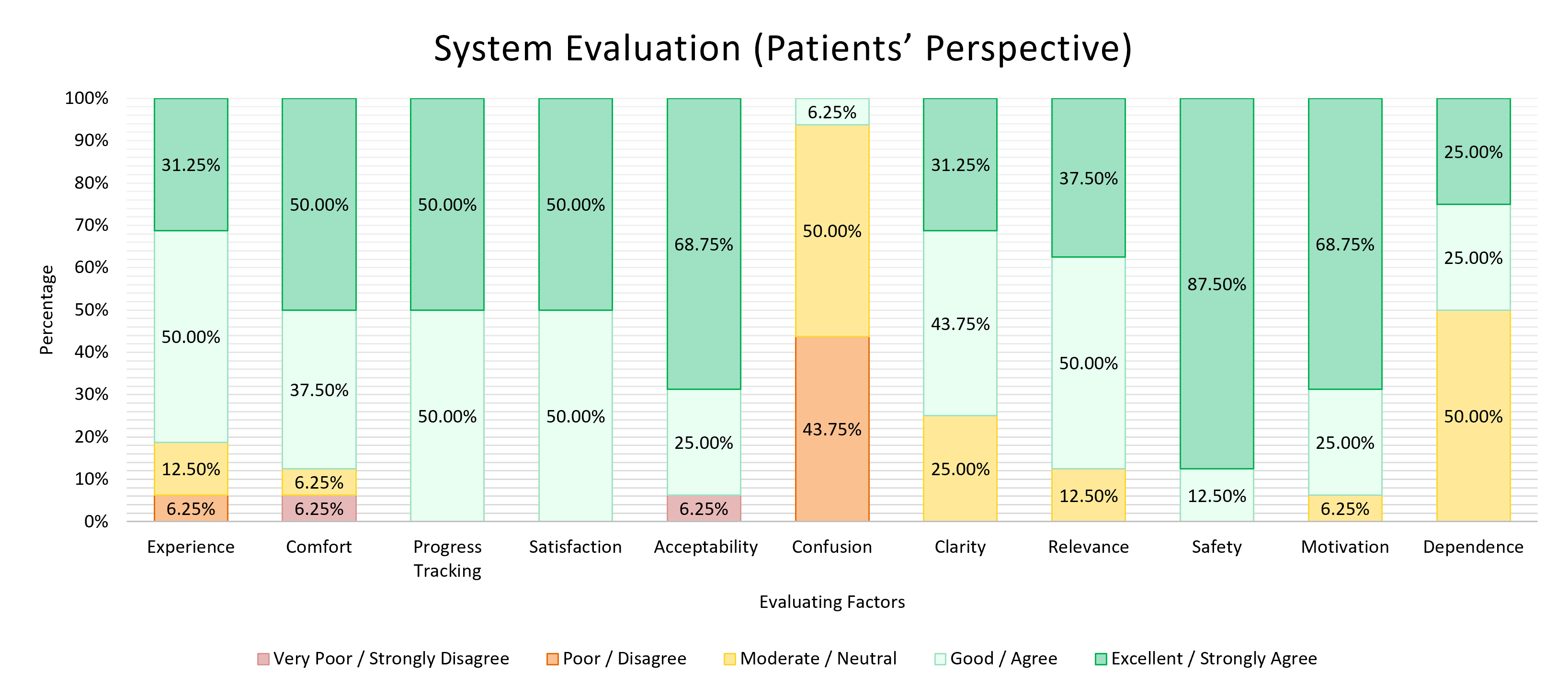}
                \caption{}
                \label{fig:useq patient}
            \end{subfigure}
            \begin{subfigure}{.95\textwidth}
                \centering
                \includegraphics[width=\textwidth]{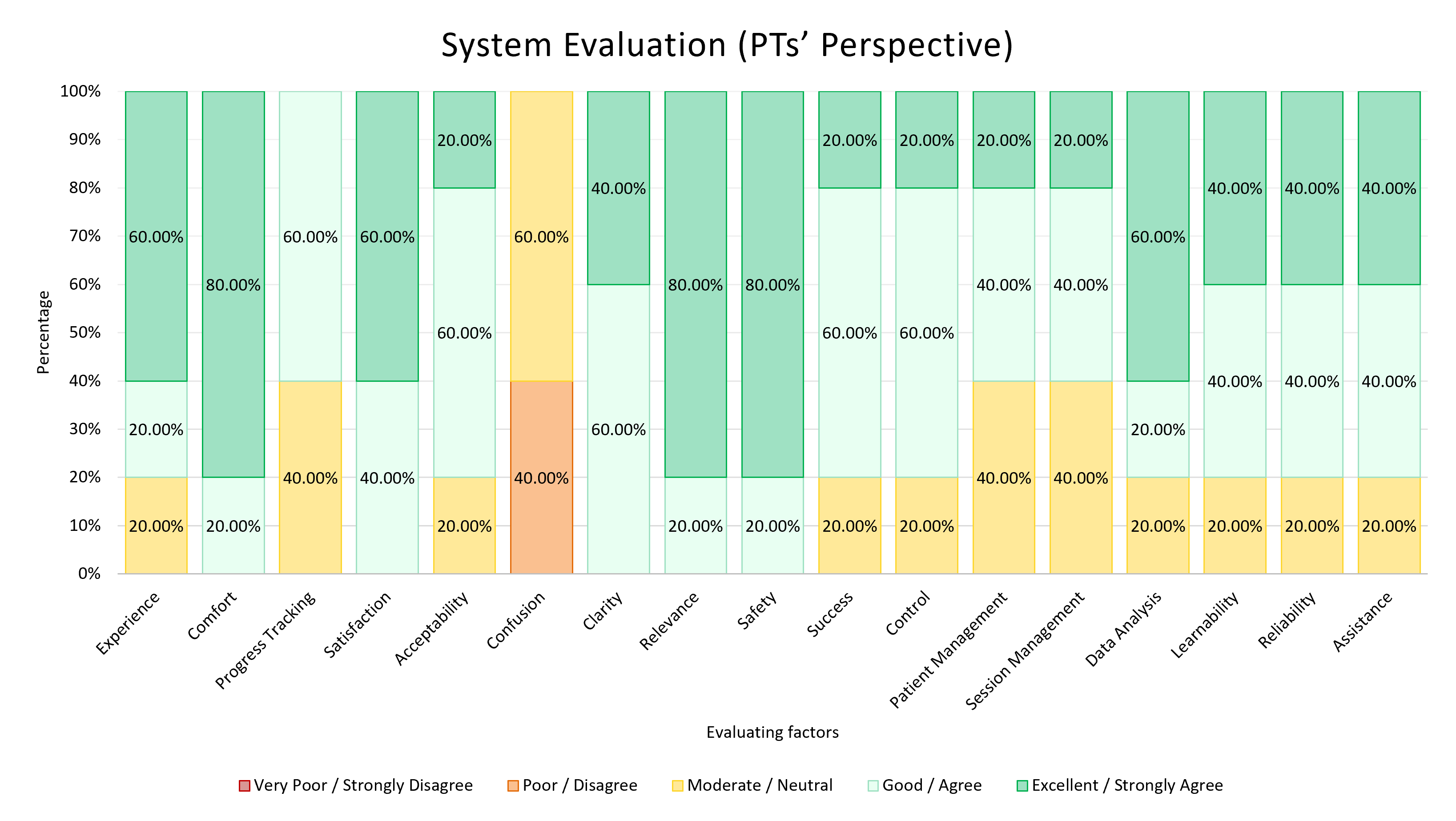}
                \caption{}
                \label{fig:useq pt}
            \end{subfigure}
            
            \caption{System evaluation outcome from the perspectives of - (a) Patients and (b) Physiotherapists.}
            \label{fig:useq patient pt}
        \end{figure}

        The means of the user ratings across all the factors, as summarized in \autoref{tab:useq rating summary}, provides substantial evidence of high user satisfaction with the system. From a general perspective, the low mean user rating for the factor \textit{confusion} ($<$3) indicates that the features and functionalities of \textit{Renovo} were less confusing. However, the physiotherapists rated the system moderately (Mean: 3.80) for the factors \textit{patient management} and \textit{session management}, indicating a scope of upgrading the system. From the perspective of the patients, the system should motivate them toward therapeutic intervention without compromising their dependence on the physiotherapists. This is confirmed by the high rating for the factor \textit{motivation} (Mean: 4.63). However, the patient rating for the factor \textit{dependence} (Mean: 3.75) suggests that they moderately depended on \textit{Renovo}. This is explainable, as \textit{Renovo} allows the patients to visualize their performance in real-time, which was previously not possible in the conventional mode of therapy. Graphical illustrations showing the ratings of the patients and the physiotherapists on their satisfaction with \textit{Renovo} are shown in \autoref{fig:useq patient} and \autoref{fig:useq pt}, respectively. 
        
\begin{figure}[htbp]
        \centering
        \begin{subfigure}{\textwidth}
            \centering
            \includegraphics[width=\textwidth]{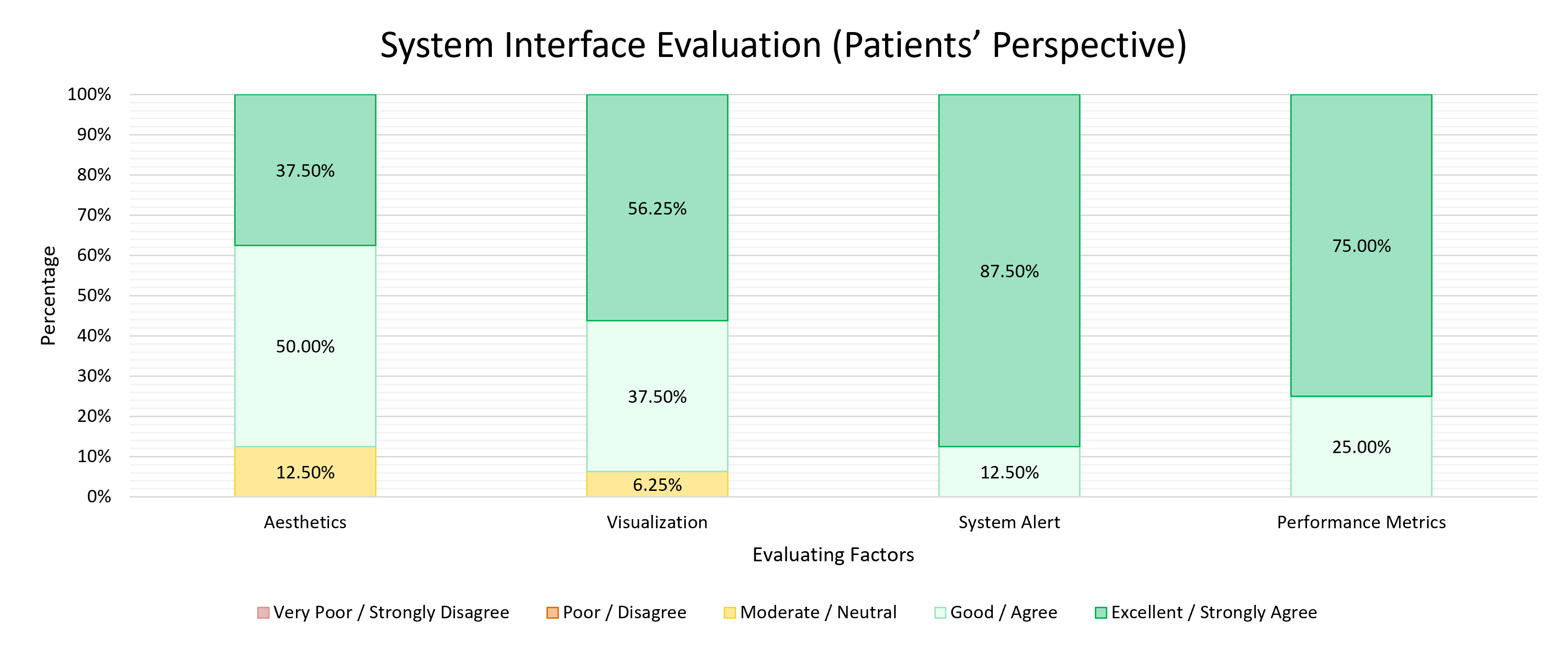}
            \caption{}
            \label{fig:quis patient}
        \end{subfigure}
        \hfill
        \begin{subfigure}{\textwidth}
            \centering
            \includegraphics[width=\textwidth]{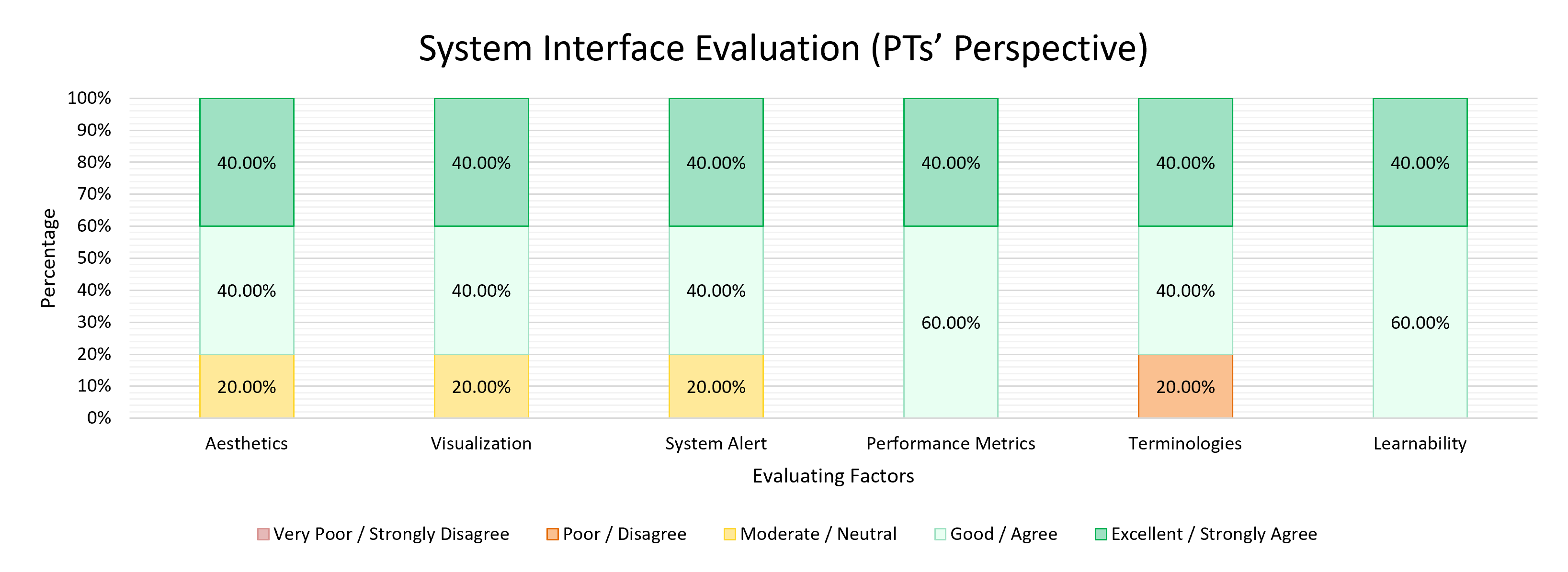}
            \caption{}
            \label{fig:quis pt}
        \end{subfigure}
        
        \caption{System interface evaluation outcome from the perspectives of - (a) Patients and (b) Physiotherapists.}
        \label{fig:quis patient pt}
    \end{figure}

        All the common factors of user satisfaction with the system interface, such as - \textit{aesthetics}, \textit{visualization}, \textit{system alert}, and \textit{performance metrics}, were highly rated by both the physiotherapists and the patients. The disjoint factors of system interface evaluation from the perspective of the physiotherapists, such as - \textit{terminologies} and \textit{learnability}, were evaluated similarly. These ratings indicate that the users of \textit{Renovo} (physiotherapists and patients) were highly satisfied with what it had to offer in the domain of rehabilitation engineering. A summary of the mean ratings of user satisfaction with the system interface, considering various common and disjoint factors, is summarized in \autoref{tab:quis rating summary}, followed by a graphical illustration in \autoref{fig:quis patient pt}. 
        
        \begin{table}[htbp]
            \centering
        	\caption{Mean ratings of user satisfaction with the user interface of \textit{Renovo}.}
        	\label{tab:quis rating summary}
        	\tiny
        	\begin{tabularx}{.9\textwidth}{ccc}
        		\toprule
        		\multicolumn{1}{c}{\multirow{2}{*}{\textbf{Factors}}}  & \multicolumn{2}{c}{\textbf{Mean Ratings}* (\textit{out of 5.00})} \\ \cline{2-3}\\
                & \textbf{Physiotherapists} & \textbf{Patients}\\
        		\midrule
                Aesthetics &4.20 &4.25\\
                Visualization &4.20 &4.50\\
                System Alert &4.20 &4.88\\
                Performance Metrics &4.40 &4.75\\
                Terminologies$ ^{dt}$ &4.00 &-\\
                Learnability$ ^{dt}$ &4.40 &-\\
                \midrule
                \multicolumn{2}{r}{\textbf{Mean rating across all the factors by Physiotherapists}} &\multicolumn{1}{c}{4.23}\\
                \multicolumn{2}{r}{\textbf{Mean rating across all the factors by Patients}} &\multicolumn{1}{c}{4.5}\\
                \bottomrule\\
                \multicolumn{3}{l}{* Ratings that are marked by a '-', are irrelevant to a particular user (physiotherapist or patient).}\\
                \multicolumn{3}{l}{$ ^{dt}$ Factors relevant to the physiotherapist only.}
        	\end{tabularx}
        \end{table}

\newpage 

\section{Discussion}

    \subsection{Research Challenges}
    In the context of this study, it was challenging for us to help the physiotherapists realize the contribution of \textit{Renovo} in improving the efficiency and effectiveness of physiotherapy while overcoming the shortcomings of the conventional method. Furthermore, ensuring the use of appropriate and widely acknowledged terminologies in the domain of rehabilitation engineering, so that the physiotherapists can use the system without any difficulty or confusion, was a challenging task as well. 
    
    In the domain of rehabilitation engineering, one of the vital constraints of developing any assistive technology, as identified by interviewing certified domain experts, is that physiotherapists should be aware of patients’ progress and be in control throughout the entire phase of rehabilitation so that proper interventions can be prescribed. Thus, it was a deliberate decision on our part not to provide any animated illustrations of any particular intervention in \textit{Renovo} so that the patients are forced to receive assistance from the physiotherapists on how to perform that intervention, leveraging the \textit{Passive} mode of therapy in the \textit{user interface}.    
    
    \subsection{Limitations and Future Works}
    Apart from motor impairment(s) of the upper limb(s), stroke patients also suffer from simultaneous motor impairments in other parts of their body as well \cite{fries1993motor}, often forcing them to be bedridden. One of the major limitations of \textit{Renovo} is that all the interventions featured in this work need to be started from a corresponding base arm posture, as illustrated in the \textit{user interface} in \autoref{fig:main interface}. If this posture is not maintained during device calibration, the system will give erroneous measurements. This seriously limits the target patients of \textit{Renovo} to only those with the ability to maintain a sitting posture during therapeutic sessions. Another limitation of \textit{Renovo} is the lack of any interventions related to the rehabilitation of finger strength and movements. In the wake of COVID-19, it became quite impossible for stroke patients to physically attend a therapeutic session, out of fear of exposure to the virus. In this regard, both the patients and the physiotherapists could benefit from a telerehabilitation feature, in terms of continuation of rehabilitation despite inconvenient and unexpected situations, which is currently unavailable in \textit{Renovo}. Another limitation of \textit{Renovo} is that physiotherapists cannot create any patient-specific therapeutic regimes. They can only monitor and track one intervention at a time, selected before a session from the list of interventions, featured in \textit{Renovo}. The impact of this can be observed from the comparatively low Progress Tracking ($3.6$), Patient Management ($3.8$), and Session Management ($3.8$) scores, as seen from \autoref{tab:useq rating summary}. About the limitations of \textit{Renovo} stated above, we are actively working to make the system independent of the posture of the patients, to incorporate the rehabilitation of motor impairments of the fingers due to stroke, to facilitate telerehabilitation, and to create patient-specific therapeutic regimes, among others, in the future. Among these, telerehabilitation and the creation of patient-specific therapeutic regimes can be incorporated by either developing our patient management system from scratch or by connecting the application with any already existing commercially available solution. However, to provide rehabilitation of motor impairments of the fingers, a different sensor setup may be required, which requires further rigorous research work. To make the system independent of the posture of the patient is another challenging task. A possible solution might be to collect a possible list of starting postures from the physiotherapists and provide those as options within the software. Alternatively, a visual creation system might be offered to the therapists so that they can customize initial postures using some simple steps.

\section{Conclusion}
    In this work, we have presented the design, development, and results of a three-week pilot study of the prototype of a wearable sensor-based therapeutic system, \textit{Renovo}, whose primary objective is to assist the physiotherapists with real-time performance visualization and evaluation of stroke patients with motor impairment(s) of the upper limb(s) during rehabilitation. Although, as a pilot study, we have obtained a proof of feasibility of \textit{Renovo} involving a limited number of patients and physiotherapists, further investigation is required to realize its full potential to be occupationally adopted by the physiotherapists as an assistive technology, making it imperative to conduct a cohort study in the future. To conclude, affordability and accessibility to therapeutic systems, such as \textit{Renovo}, may assist physiotherapists with their evaluation of patients' progress, and at the same time, increase the efficiency and effectiveness of the rehabilitation of stroke patients with motor impairment(s) of the upper limb(s).

\section*{Acknowledgements}
    This research was not funded by any private or public agencies. The authors express their heartfelt gratitude to the participants for their valuable time and effort in making this study possible. 

\section*{Conflict of Interest}
    The authors do not declare any conflict of interest that may alter the outcomes of the study in any manner, and approve this version of the manuscript for publication.




 \bibliographystyle{elsarticle-num} 
 \bibliography{Renovo_Citations}





\end{document}